\documentclass[iop]{emulateapj}
\usepackage{amsmath}
\usepackage{hyperref}
\usepackage{color}
\usepackage{array}
\usepackage{listings}
\usepackage{multirow}
\usepackage{sidecap}

\begin{document}

\title{On magnetic equilibria in barotropic stars}

\author{Crist\'obal Armaza\altaffilmark{1}, Andreas Reisenegger\altaffilmark{1}, and Juan Alejandro Valdivia\altaffilmark{2}}

\altaffiltext{1}{Instituto de Astrof\'isica, Pontificia Universidad Cat\'olica de Chile, Av. Vicu{\~n}a Mackenna 4860, 782-0436 Macul, Santiago, Chile. \email{cyarmaza@uc.cl}}

\altaffiltext{2}{Departamento de F\'isica, Facultad de Ciencias, Universidad de Chile, Casilla 653, Santiago, Chile.}





\begin{abstract}
Upper main sequence stars, white dwarfs and neutron stars are known to possess stable, large-scale magnetic fields. Numerical works have confirmed that stable MHD equilibria can exist in non-barotropic, stably stratified stars. On the other hand, it is unclear whether stable equilibria are possible in barotropic stars, although the existing evidence suggests that they are all unstable. This work aims to construct barotropic equilibria in order to study their properties, as a first step to test their stability. We have assumed that the star is a perfectly conducting, axially symmetric fluid, allowing for both poloidal and toroidal components of the magnetic field. In addition, we made the astrophysically justified assumption that the magnetic force has a negligible influence on the fluid structure, in which case the equilibrium is governed by the Grad--Shafranov equation, involving two arbitrary functions of the poloidal flux. We built a numerical code to solve this equation, allowing for an arbitrary prescription for these functions. Taking particularly simple, but physically reasonable choices for these functions with a couple of adjustable parameters, all of the equilibria found present only a small ($\lesssim 10\%{}$) fraction of the magnetic energy stored in the toroidal component, confirming previous results. We developed an analytical model in order to study in more detail the behavior of the magnetic energy over the full range of parameters. The model confirms that the toroidal fraction of the energy and the ratio of toroidal to poloidal flux  are bounded from above for the whole range of parameters.
\end{abstract}

\begin{keywords}
 {magnetohydrodynamics (MHD); stars: magnetic field; stars: neutron; stars: white dwarfs}
\end{keywords}

\section{Introduction}
\label{sec:intro}

In contrast to the magnetic fields of low-mass stars (e.g. the Sun), magnetic fields in upper main sequence stars, white dwarfs, and neutron stars are quite stable on time scales comparable to their life times. However, not all of these stars are actually ``magnetic,'' in the sense that massive stars and white dwarfs follow a bimodal distribution in which only a fraction within $5-20\,$\%{} of each star type stands out for having strong magnetic fields when compared to the remaining $90\%$ \citep{liebert3, ferrario6, auriere7, kawka7, donati9, wade14}. Throughout this work, we will be interested in the stable fields in upper main sequence stars, white dwarfs, and neutron stars. 

From a theoretical point of view, the persistence of magnetic fields in massive stars and compact remnants motivates the interest in what physical conditions are involved in sustaining such equilibria. Both analytical and numerical studies have generally pointed out that stable equilibrium configurations require a poloidal component of the magnetic field to stabilize a toroidal one, and vice versa \citep{prendergast56, braithwaite4, braithwaite6, akgun13}, while purely poloidal and purely toroidal equilibria undergo intrinsic instabilities related to their geometries \citep{markey73, tayler73, wright73, flowers77, kiuchi8, ciolfi11, lasky11, marchant11}. The simulations of Braithwaite and collaborators showed that initially random fields often evolve naturally into nearly axisymmetric, toroidal-poloidal ``twisted-torus'' configurations on short (Alfv\'en) time scales. 

In addition, an important ingredient to consider is that the matter within the stars considered is non-barotropic, that is, the pressure is a function of mass density and another thermodynamic quantity, which is conserved in a fluid element if the latter is perturbed from its equilibrium position \citep{reisenegger9}. This second thermodynamic variable accounts for stable stratification, i.e., stability against convective motion. On short timescales, massive stars and white dwarfs are stratified by entropy gra\-dients, while in neutron stars this role is played by a varying chemical composition. Stable stratification could be a crucial ingredient in stabilizing these equilibria, inhibiting radial motions and hence suppressing fluid instabilities occurring within these stars, as supported by recent numerical simulations \citep{mitchell14}. As a matter of fact, it has recently been suggested that a plausible condition of dynamical stability for these magnetized stars can be expressed as 
\begin{equation}\label{rpol}
2a(E_\text{mag}/E_\text{grav}) < E_\text{pol}/E_\text{mag} \lesssim 0.8,
\end{equation} 
where $E_\text{mag}/E_\text{grav}$ is the magnetic-to-gravitational energy ratio, $E_\text{pol}/E_\text{mag}$ is the fraction of magnetic energy stored in the poloidal component, and $a$ is a pure number accounting for how stratified the star is \citep{braithwaite9, akgun13}. The additional fact that all of these objects are expected to contain a gravitational energy much larger than the magnetic energy \citep{reisenegger9} would imply that the toroidal component can be stabilized by a much weaker poloidal field, but the converse is not true. This is of particular relevance in the context of energy released by magnetars, where it has been claimed that a much stronger internal toroidal field would account for observed outbursts emitted from these objects \citep{thompson96}. 

Stable stratification, although crucial on short time scales, could be eroded if some processes change either entropy or chemical composition. In massive stars and white dwarfs, heat conduction is too slow to significantly change entropy during their life time. For neutron stars, the stable stratification could be overcome by changing the chemical composition by means of $\beta$-decay and/or ambipolar difusion \citep{goldreich92, pethick92, reisenegger9}, making them effectively barotropic over long time scales.

On the other hand, and despite their lack of realism, barotropic equations of state have often been assumed to describe the matter within upper main sequence stars, white dwarfs and neutron stars \citep{yoshida6, haskell8, lander9, ciolfi9, fujisawa12, pili14, bucciantini15}, because of the expected simplicity as a very first approximation to the realistic state of matter. Yet, it is important to identify correctly when one is actually modeling a barotropic fluid. The local adiabatic index of the fluid, $\Gamma = (\partial\ln P/\partial\ln \rho)_X$, where $X$ is the specific entropy (or another quantity conserved in fast fluid displacements), will not generally be equal to the analogous index describing the equilibrium, $\gamma = d\ln P/ d\ln\rho$. If $\Gamma = \gamma$, we say that the fluid is barotropic.  A single-species, non-interacting degenerate Fermi gas, for example, \emph{does} possess a barotropic equation of state, which in limiting cases of nonrelativistic and ultrarelativistic particles reduces to polytropes with $\Gamma = \gamma = 5/3$ and $\Gamma = \gamma = 4/3$, respectively. 

Barotropy strongly restricts the allowed range of possible equilibrium configurations and, as discussed, does not strictly represent the realistic stably stratified matter within these objects on short (Alfv\'en) time scales. Moreover, the question whether magnetic equilibria in barotropic stars can be stable or not is being answered negatively by recent studies \citep{lander12, mitchell14}. 

Several authors \citep{ciolfi9, lander9, fujisawa12, gourgouliatos13, pili14, bucciantini15} have explored the possible axially symmetric equilibria in barotropic stars, generally finding that the fraction of the total magnetic energy corresponding to the toroidal component, $E_\text{tor}/E_\text{mag}$, is at most a few \%{}, even in cases of comparable poloidal and toroidal magnetic field strength. Since stable stratification, which is absent in the barotropic case, is expected to be a key piece in the stability of the stars considered here, and given such a small fraction of toroidal energy apparently inherent to barotropic configurations, it is likely that all of these equilibria are dynamically unstable, as supported by recent simulations \citep{lander12, mitchell14}. More recent works \citep{ciolfi13, fujisawa13} have shown, however, that higher fractions $E_\text{tor}/E_\text{mag}$ are actually possible, making a more extensive survey of these equilibria relevant. Studying properties of barotropic equilibria could also be relevant considering the scenario in which neutron stars would reach an effectively barotropic state after overcoming stable stratification by means of dissipative processes, as already discussed.

The present work is focused on obtaining a wide range of barotropic equilibria, paying attention to their main properties. In addition, these results may be considered as a starting point to study in more detail whether magnetic fields in barotropic stars can be stable or not. This paper is organized as follows. \S 2 presents the formalism used to construct barotropic equilibria, in order to introduce the notation used throughout this paper. In \S 3 we solve numerically the resulting equilibrium equation. We summarize the tests carried out to check our code and present the main results obtained. \S 4  expands the analysis of barotropic equilibria, introducing an asymptotic, analytical model to explore equilibrium configurations beyond numerical simulations. Finally, \S 5 presents the conclusions of this paper.

\section{Magnetic equilibria in barotropic stars}

We assumed that the star is a perfectly conducting fluid. In the ideal MHD approximation, the dynamical equilibrium of a star is described by the Euler equation,
\begin{equation}
0 = -\boldsymbol\nabla P - \rho \boldsymbol\nabla\Psi + \frac{1}{c}\mathbf J\times \mathbf B,  \label{euler_mag}
\end{equation}
where $P$, $\rho$, $\Psi$, $\mathbf B$, $\mathbf J$ and $c$ are the fluid pressure, mass density, gravitational potential, magnetic field, electric current density and speed of light, respectively, so that the last term in Eq. \eqref{euler_mag} is the Lorentz force per unit volume. Throughout this work, a spherical coordinate system $(r,\theta,\phi)$ with origin at the center of the star is used to describe all quantities. Also, we assumed that the current density drops toward the stellar surface and vanishes outside, as expected since the mass density and the charged-particle density do so. For simplicity, we considered axially symmetric magnetic stars. The magnetic field may then be expressed as the sum of a poloidal and a toroidal component, each determined by a single scalar function \citep{chandra56},
\begin{equation}\label{decomp}
\mathbf B = \mathbf B_\text{pol} + \mathbf B_\text{tor} = \boldsymbol\nabla\alpha(r,\theta)\times \boldsymbol\nabla\phi + \beta(r,\theta)\boldsymbol\nabla\phi,
\end{equation}
where $\boldsymbol\nabla\phi = \mathbf e_\phi/r\sin\theta$, $\mathbf e_i$ being the orthonormal basis of the coordinate system, with $i=\{r,\theta,\phi\}$. By Amp\`ere's law, the electric current density reads 
\begin{equation}
\frac{4\pi}{c}\mathbf J = \boldsymbol\nabla\beta \times \boldsymbol\nabla\phi - \Delta^*\alpha\,\boldsymbol\nabla\phi,\label{J}
\end{equation}
where the Laplacian-like ``Grad--Shafranov operator''
\begin{align}\nonumber
\Delta^* &\equiv r^2\sin^2\theta\, \boldsymbol\nabla\cdot \left[\frac{1}{r^2\sin^2\theta}\boldsymbol\nabla\right]\\  
&= \frac{\partial^2}{\partial r^2} + \frac{\sin\theta}{r^2}\frac{\partial}{\partial \theta}\left( \frac{1}{\sin\theta}\frac{\partial}{\partial \theta} \right)\label{GSop}
\end{align}
was introduced. For an axisymmetric equilibrium, the azimuthal component of the magnetic force must vanish, which is imposed by demanding that $\beta$ be a function of $\alpha$ only, $\beta = \beta(\alpha)$. This implies that both $\alpha$ and $\beta$ are constant along poloidal field lines. Also, notice that, since the poloidal component is perpendicular to the toroidal one by construction, one can always split the magnetic energy in terms of the energy stored in each component, 
\begin{equation}
E_\text{mag} = E_\text{pol} + E_\text{tor} = \int  \frac{\mathbf B_\text{pol}^2}{8\pi}\,dV + \int \frac{\mathbf B_\text{tor}^2}{8\pi}\,dV,
\end{equation}
where the integration is carried out over all space.

\subsection{Magnetic field outside the star.}

Since we demanded vacuum outside the star, $\beta$ has to be a constant there (see Eq. \eqref{J}). This constant must be zero in order to have a finite magnetic field along the axis. Also, it is needed that 
\begin{equation}
\Delta^*\alpha=0,
\end{equation}
 which is a linear, homogeneous partial differential equation, whose general solution after separating variables is
\begin{equation}
\alpha(r,\theta) = \sum_{\ell=1}^\infty \sin\theta\left[a_\ell r^{\ell+1} + b_\ell r^{-\ell}\right]P_\ell^1(\cos\theta),\label{expoutsideGeneral}
\end{equation}
where $P_\ell^m(x)$ is the associated Legendre polynomial of order $\ell$ with azimuthal index $m$. Since the magnetic field must vanish far from the star, $a_\ell = 0$ for all values of $\ell$, and
\begin{align}\label{expoutside}
\alpha &= \sum_{\ell=1}^\infty b_\ell \frac{\sin\theta}{r^\ell}P_\ell^1(\cos\theta),\\
B_r &= \frac{1}{r^2\sin\theta}\frac{\partial\alpha}{\partial\theta} = -\sum_{\ell = 1}^\infty\frac{\ell(\ell + 1) b_\ell}{r^{\ell+2}}P^0_\ell(\cos\theta),\\
B_\theta &= -\frac{1}{r\sin\theta}\frac{\partial \alpha}{\partial r} = \sum_{\ell=1}^\infty\frac{\ell b_\ell}{r^{\ell + 2}} P_\ell^1(\cos\theta)
\end{align}
outside the star.

\subsection{Magnetic field inside the star: the GS equation.}

Throughout this work, we considered a barotropic equation of state, $P=P(\rho)$, in which case if we divide Eq. \eqref{euler_mag} by $\rho$ and take the curl of the result, the Lorentz force per unit mass must be the gradient of a certain axisymmetric scalar function $\chi(r,\theta)$, $\mathbf f_\text{mag}/\rho  = \boldsymbol\nabla\left[\chi/4\pi\right]$, where the factor $4\pi$ is introduced for convenience. Replacing the decomposition of Eq. \eqref{decomp} in the Lorentz force, it turns out that $\chi$ must be a function of $\alpha$ as well. Hence $\boldsymbol\nabla\chi = \chi'\boldsymbol\nabla\alpha$, and the so-called Grad--Shafranov equation (hereafter GS, \citealt{grad58}, \citealt{shafranov66}),
\begin{equation}
\Delta^* \alpha + \beta(\alpha)\beta'(\alpha) +  \rho r^2\sin^2\theta\chi'(\alpha) = 0,\label{grad}
\end{equation}
is obtained, where $\Delta^*$ is the Grad--Shafranov operator given by Eq. \eqref{GSop}. In the latter, a prime indicates derivative with respect to the argument. In this way, for given functions $\beta(\alpha)$ and $\chi(\alpha)$, the GS equation may be solved for the unknown $\alpha$. This can be done self-consistently with the Poisson equation (relating the gravitational potential and the density profile) and the Euler equation, provided an equation of state. Imposing appropriate boundary conditions at the surface, this procedure gives $\alpha$, which determines the magnetic field inside the star. This approach has been successfully used to get (numerical) barotropic equilibria \citep{lander9, fujisawa12}, but in what follows we will briefly discuss an additional, useful approximation which allowed us to obtain suitable equilibria with less calculations.

It is clear that the magnetic field can deform the star, so that $\rho$ is in principle affected by $\alpha$. Nevertheless, the fact that the magnetic energy stored in the stars considered is much smaller than their gravitational energy \citep{reisenegger9} suggests that this deformation is correspondingly small, so the density profile appearing in Eq. \eqref{grad} can be taken to an excellent approximation to be as in the non-magnetic case. This approximation has been already considered \citep{ciolfi9, gourgouliatos13}. One advantage of this approach is that the results can be scaled to any (small) field strength, which is not true for the ``self-consistent'' one. Thus, in this work we looked for barotropic equilibria by solving the GS equation for a \emph{given} density profile $\rho =\rho_0(r)$, having been particularly interested in comparing these two approaches.

\subsection{Variational principle}

Alternatively, the GS equation can also be obtained by extremizing the functional $I[\alpha] = \int\mathcal L\,dV$ with respect to the scalar function $\alpha$, where the Lagrangian density is
\begin{equation}
\mathcal L = \frac{1}{8\pi r^2\sin^2\theta}\left[|\boldsymbol\nabla\alpha|^2 - \beta^2(\alpha) - 2 \rho r^2\sin^2\theta \chi(\alpha) \right],
\end{equation}
subject to the condition that $\delta\alpha$ must vanish at the stellar surface, i.e., homogeneous Dirichlet boundary conditions. This expression generalizes that given by \cite{monaghan76}.

\subsection{Boundary conditions.}

Maxwell's equation $\boldsymbol\nabla \cdot\mathbf B = 0$ imposes that the radial component of the magnetic field be continuous across the surface. Also, the $\theta$ and $\phi$ components of the magnetic field must be continuous in order to avoid surface currents. In terms of $\alpha$, continuity of $B_r$ demands that $\partial\alpha/\partial\theta$ be continuous, which in turn implies that $\alpha$ itself must be continuous,
\begin{equation}\label{auxaux}
\alpha_\text{in}(R,\theta) = \alpha_\text{out}(R,\theta) = \sum_{\ell=1}^\infty \frac{b_\ell}{R^\ell}\sin\theta P^1_\ell(\cos\theta),
\end{equation}
where we have introduced the expansion for $\alpha$ outside, while $\alpha_\text{in}(r,\theta)$ is a solution of the GS equation inside the star. Using the completeness of the set $\{P_\ell^m(x)\}$, one can extract the coefficients $b_\ell$ appearing in Eq. \eqref{auxaux} in terms of the value of $\alpha$ at the surface, leading to
\begin{equation}
b_\ell =  R^\ell\frac{2\ell + 1}{2\ell(\ell + 1)}I_\ell,\label{aell}
\end{equation}
where $I_\ell$ stands for the integral
\begin{equation}\label{Iell}
I_\ell \equiv \int_0^\pi P^1_\ell(\cos\theta)\alpha(R,\theta)\,d\theta.
\end{equation}
This choice of the coefficients $b_\ell$ ensures $B_r$ continuous. Continuity $B_\theta$ implies that
\begin{align}\nonumber
\left.\frac{\partial \alpha_\text{in}}{\partial r}\right|_{r=R} &= \left.\frac{\partial \alpha_\text{out}}{\partial r}\right|_{r=R} = -\frac{1}{R}\sum_{\ell=1}^\infty\frac{\ell b_\ell}{R^\ell}\sin\theta P^1_\ell(\cos\theta)\\
 &= -\frac{1}{R}\sum_{\ell=1}^\infty\frac{2\ell + 1}{2\ell + 2}\sin\theta P^1_\ell(\cos\theta)I_\ell,\label{dadr}
\end{align}
where in the last equality we have replaced the expression for $b_\ell$ found in Eq. \eqref{aell}. Finally, in order to have $B_\phi$ continuous, we demanded that $\beta = 0$ at the surface, as explained in more detail in the next subsection.

\subsection{Particular choices for the magnetic functions.}

As mentioned, $\beta(\alpha)$ is an arbitrary function of $\alpha$ only. However, its shape is physically constrained as follows. Since $\beta = 0$ outside and $\beta$ is constant along poloidal field lines, it has to be zero along poloidal field lines closing outside the star. That is, the toroidal field may lie only in regions where the poloidal field lines close \emph{inside} the star. A particularly simple choice to achieve this is to set
\begin{equation}\label{twisted}
\beta(\alpha) = \begin{cases}
s(\alpha - \alpha_\text{surf})^\lambda & \text{if}\quad \alpha \geq \alpha_\text{surf}\\
0 & \text{if}\quad \alpha < \alpha_\text{surf},
\end{cases}
\end{equation}
where $s$ and $\lambda$ are free parameters, while $\alpha_\text{surf} \equiv \alpha(R,\pi/2)$ is the value of $\alpha(r,\theta)$ at the surface of the star on the equator, which is also the value of $\alpha$ along the longest poloidal field line closing inside the star. This choice has already been used by previous authors \citep{ciolfi9, lander9, fujisawa12, lander12, akgun13, gourgouliatos13}. It is fundamental to notice that geometrically one does not know a priori where to locate the boundary between the volume with and without toroidal field. Also, $\alpha_\text{surf}$ is \emph{not} a given parameter, but it has to be determined while solving the GS equation. We set $\lambda = 1.1$ in order to compare with previous works. It is expected that higher values of $\lambda$ result in smaller contributions of the $\beta\beta'$ term in the GS equation, leading to a smaller toroidal field strength, although it turns out that there are no significant differences in the configurations when using larger $\lambda$. The parameter $s$, instead, plays a significant role in the configurations obtained, so a separate subsection will be devoted to this dependence. Throughout this work, we also chose
\begin{equation}
\chi'(\alpha) = k\quad\text{(constant)}.
\end{equation}

\section{Numerical tests and results}

We developed a finite-difference code to solve the GS equation for arbitrary choices of the functions $\beta(\alpha)$, $\chi'(\alpha)$, and $\rho_0(r)$. Our code solves for $\alpha$ on a polar grid of $N_r$ points in the $r$-direction and $N_\theta$ points in the $\theta$-direction, corresponding to the inside of the star, and it matches the solution smoothly to the multipolar expansion of Eq. \eqref{expoutside}. Since the infinite sum defining the multipolar expansion outside cannot be performed numerically, we truncate it to a maximum multipole $\ell_\text{max}$, defined at the beginning of the method. The details of our code are summarized in the Appendix. 

We solved the GS equation for different density profiles, considering the particular function $\beta(\alpha)$ given in Eq. \eqref{twisted} and normalized our results so that $\chi'(\alpha) = k = 1$. We also normalized distances to stellar radius and density profiles to $\rho(0) = 1$. The results can be easily scaled to other normalizations.

\subsection{Tests performed to our code.}

After obtaining numerical solutions, we tested whether they are actually solutions by finding that, for resolutions $N_r = 200$ and $N_\theta = 300$, the inequality 
\begin{equation}
|\Delta^*\alpha + \beta'\beta + \rho \chi' r^2\sin^2\theta| \leq \varepsilon \left|\alpha_\text{max}\right|
\end{equation}
held for $|\varepsilon| = 10^{-8}$ on each point on the grid. Here $\alpha_\text{max}$ is the maximum value of $\alpha$ reached by the corresponding equilibrium over the space. This inequality held for all equilibria obtained with this resolution, while better accuracy was obtained for higher resolutions, as expected. Regarding boundary conditions, we confirmed that the coefficients that define $\alpha$ outside the star were consistently calculated by getting that the inequality
\begin{equation}
\alpha_\text{in}/\alpha_\text{out} \leq 1 + \varepsilon
\end{equation}
held on each point on the surface, with $|\varepsilon| = 10^{-6}$ for $N_r = 200$, $N_\theta = 300$. We also tested that the radial derivative of $\alpha$ remained continuous by obtaining that the inequality 
\begin{equation}
\left|\left(\frac{\partial \alpha_\text{in}}{\partial r}\right)_{r=1} - \left(\frac{\partial \alpha_\text{out}}{\partial r}\right)_{r=1}\right| \leq \varepsilon \left|\alpha_\text{max}\right|
\end{equation}
held for $|\varepsilon| = 10^{-8}$ on each point on the surface, again, for $N_r = 200$ and $N_\theta = 300$. For the two latter tests, better resolutions improved the accuracy, but large ($>60$) values of $s$ in the prescription of $\beta(\alpha)$ yielded values of $|\varepsilon| \geq 10^{-3}$, in both cases.

Regarding the number of multipoles outside the star, one could think that, since the multipolar expansions extend to $\ell = \infty$, the larger $\ell_\text{max}$, the more accuracy one gets. However, for a fixed resolution, we found that the accuracy when calculating the multipolar coefficients drops off after passing a certain critical value of $\ell_\text{max}$.  Since we expect that the equilibria we are looking for have higher multipoles, we are interested in finding an optimal $\ell_\text{max}$. After repeating the previous tests for different resolutions, we found that $\ell_\text{max} \sim 0.1 N_\theta$ gives the best results, so we fixed this parameter to that value. 

We tested our code against known analytical solutions of the GS equation, corresponding to purely poloidal fields (for which $\beta = 0$ identically). Here we present a test using the purely poloidal field with \citep{gourgouliatos13}
\begin{equation}
\frac{\alpha(r,\theta)}{B_\text{p}R^2} = \begin{cases}
\left[\dfrac{35x^2}{16}-\dfrac{21x^4}{8} + \dfrac{15x^6}{16}\right]\sin^2\theta, & x < 1\\
\dfrac{\sin^2\theta}{2x}, & x > 1,
\end{cases}
\end{equation}
where $x \equiv r/R$ and  $B_\text{p}$ is the magnetic field at the pole. This configuration is obtained by choosing $\rho_0(r) = \rho_c(1 - r^2/R^2)$, where $\rho_c$ is the central density. For this case, we found that
\begin{equation}
\left|\alpha_\text{num} - \alpha_\text{anal}\right| \leq \varepsilon\, |\alpha_\text{max} |
\end{equation}
with $|\varepsilon| = 2\times 10^{-5}$ on each point on the grid, for $N_r = 200$ and $N_\theta = 300$. This excellent agreement increased the confidence in our numerical scheme.

We compared some of our equilibria with those obtained by \cite{gourgouliatos13} in the context of Hall equilibria in neutron star crusts. In terms of physics, the derivation and the physical interpretation of some quantities involved are quite different, but the mathematical form of the equation is exactly the same. Gourgouliatos' code is based on a Gauss--Seidel method on a cylindrical grid extending outside the star as well, and assuming as exterior boundary condition a dipolar field at the boundary of the grid. Table \ref{withkostas} summarizes this comparison, in which we display the toroidal-to-poloidal energy ratio for different values of $s$. As seen, both codes agree for all equilibria tested.
\begin{table}
\begin{center}
\begin{tabular}{|c||cccccc|}
\hline
	&	$s = 5$	&	10	&	15	&	20	&	25	&	30	\\
\hline
\hline
Gourgouliatos	&	0.15	&	0.60	&	1.4	&	2.3	&	3.2	&	3.9	\\
this work	&	0.14	&	0.57	&	1.3	&	2.2	&	3.1	&	3.7	\\
\hline
\end{tabular}
\end{center}
\caption{Fraction of energy (in \%{}) stored in the toroidal component of the magnetic equilibria obtained by the code of K. N. Gourgouliatos \citep{gourgouliatos13} vs. those obtained in this work, using the twisted-torus configuration given by Eq. \eqref{twisted}, and assuming $\rho_0(r) = \rho_c(1-r^2/R^2)$ and $k = 1$, for $N_r = 200$ and $N_\theta = 300$ in our code.}\label{withkostas}
\end{table}

\subsection{Magnetic polytropes}

Keeping the same form of $\beta(\alpha)$ given in Eq. \eqref{twisted}, we also found configurations using density profiles derived from non-magnetic equilibria with a polytropic relation between pressure and density, 
\begin{equation}
P_0(\rho_0(r)) = K[\rho_0(r)]^{1+1/n},
\end{equation}
where $K$ is a constant depending on physical parameters of the star, and $n$ is the polytropic index. As known, polytropes of index $n \geq 5.0$ are not of physical interest as the star extends to infinity. Moreover, non-magnetic polytropes with $n > 3.0$ are known to be unstable against radial perturbations (see for example \citealt{shapiro83}). Thus we restricted ourselves to obtain configurations up to $n = 3.0$, although, as said before, the code accepts any density profile. 

\begin{figure}
\begin{center},
\includegraphics[width=6.5cm]{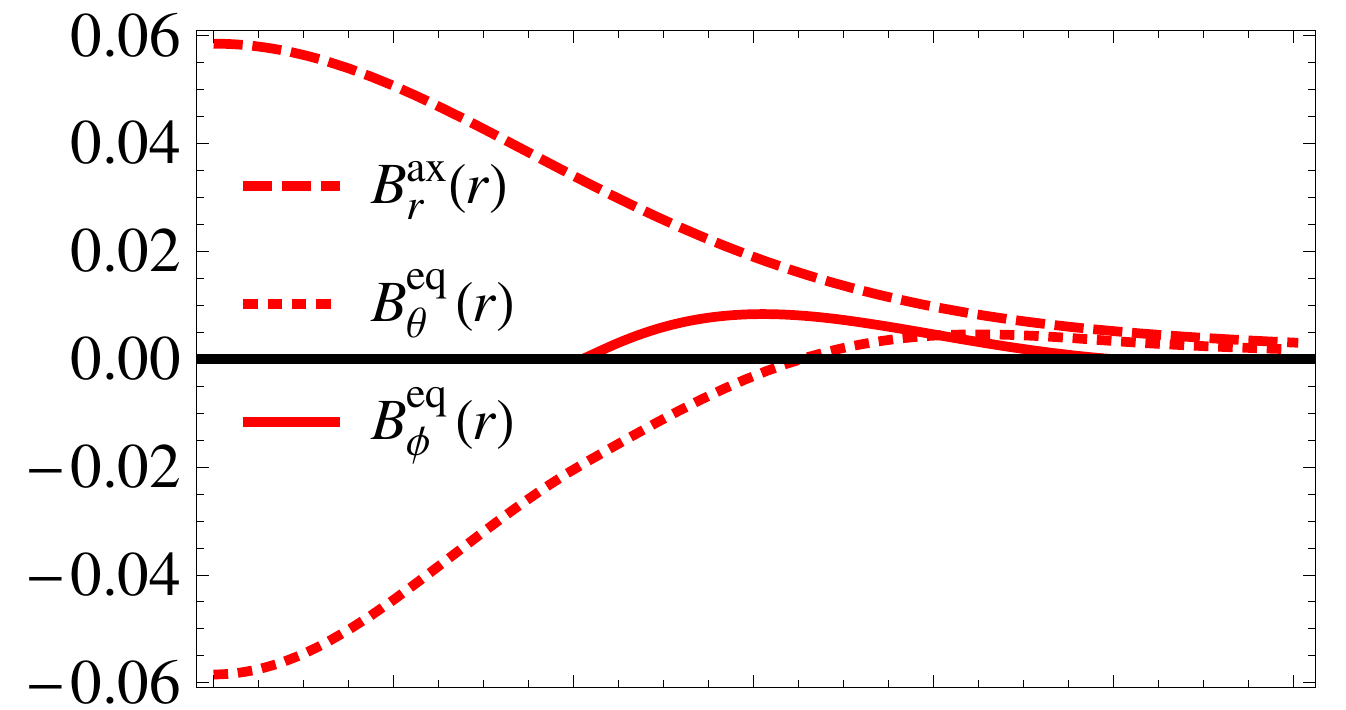}\\\vspace{0.2cm}
\includegraphics[width=6.5cm]{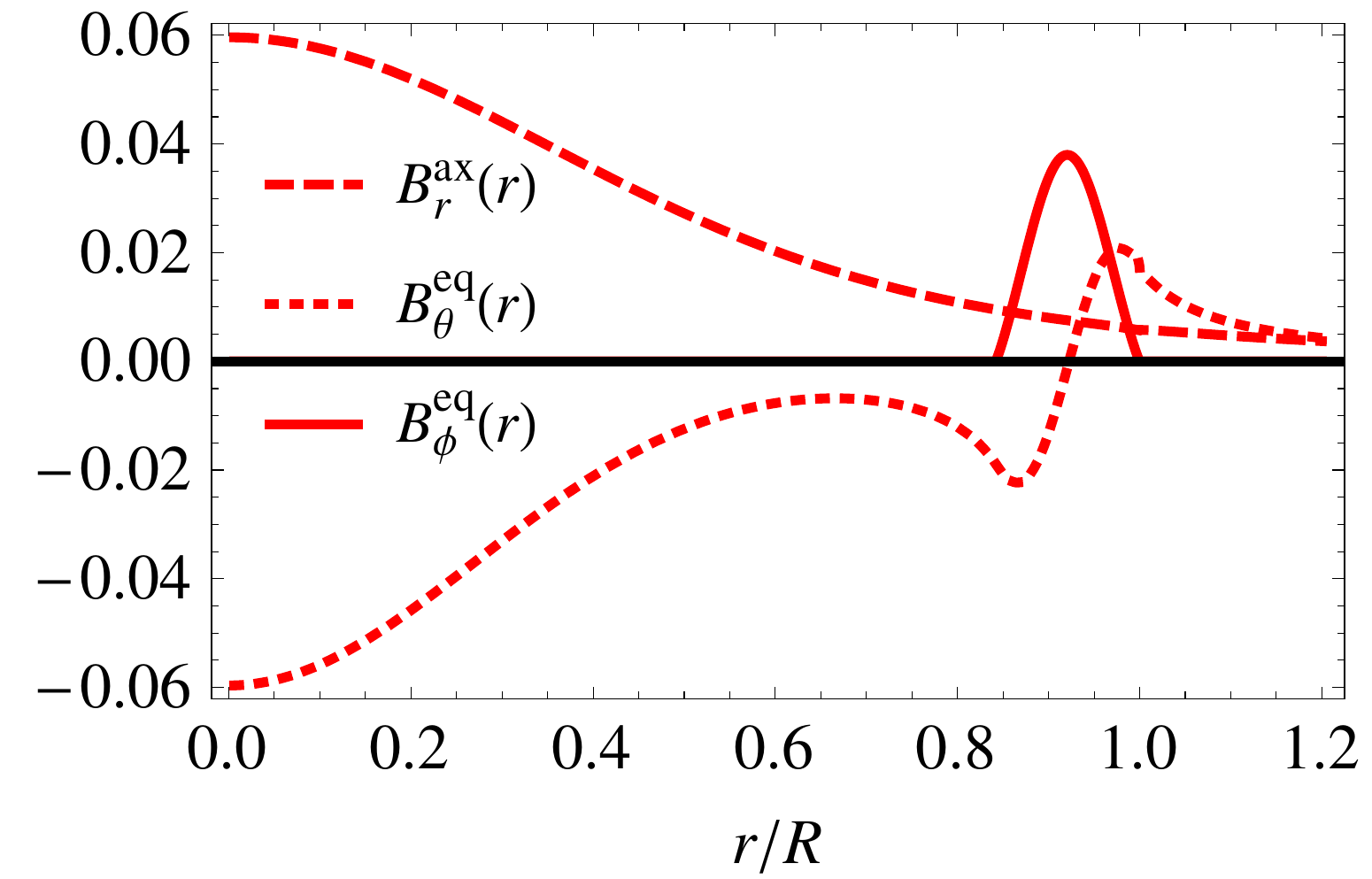}\\
\caption{Magnetic field  along the axis and the equator for $n=2$ polytropes, assuming $\chi'(\alpha) = 1$ and $\beta(\alpha)$ given in Eq. \eqref{twisted}. Top: $s = 10$. Bottom: $s = 60$. For these cases, we took $N_r = 400$ and $N_\theta = 500$.}\label{magpolyt1}
\end{center}
\end{figure}

\begin{figure*}
\begin{center}
\includegraphics[width=5.5cm]{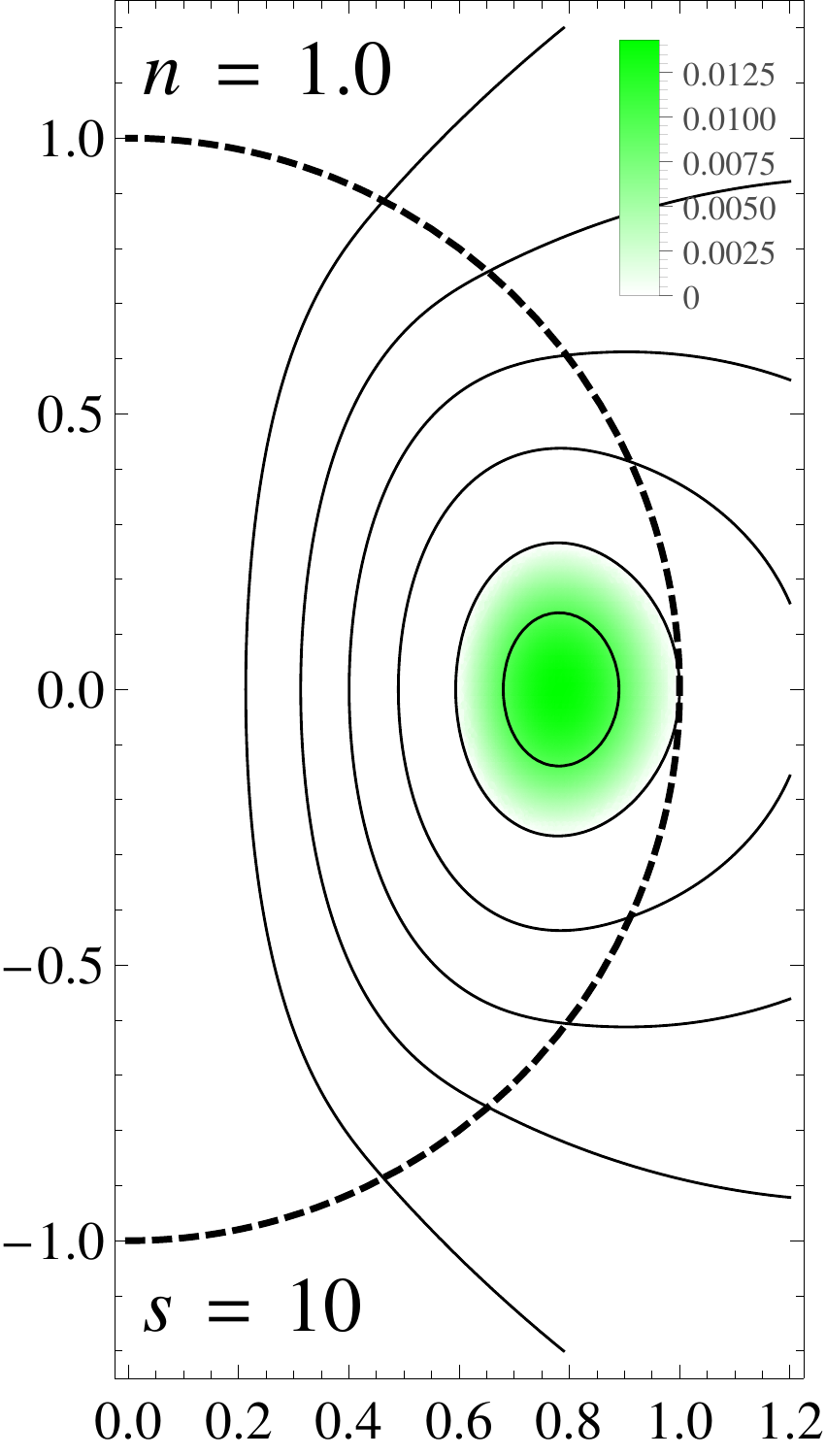}\hspace{1cm}\includegraphics[width=5.5cm]{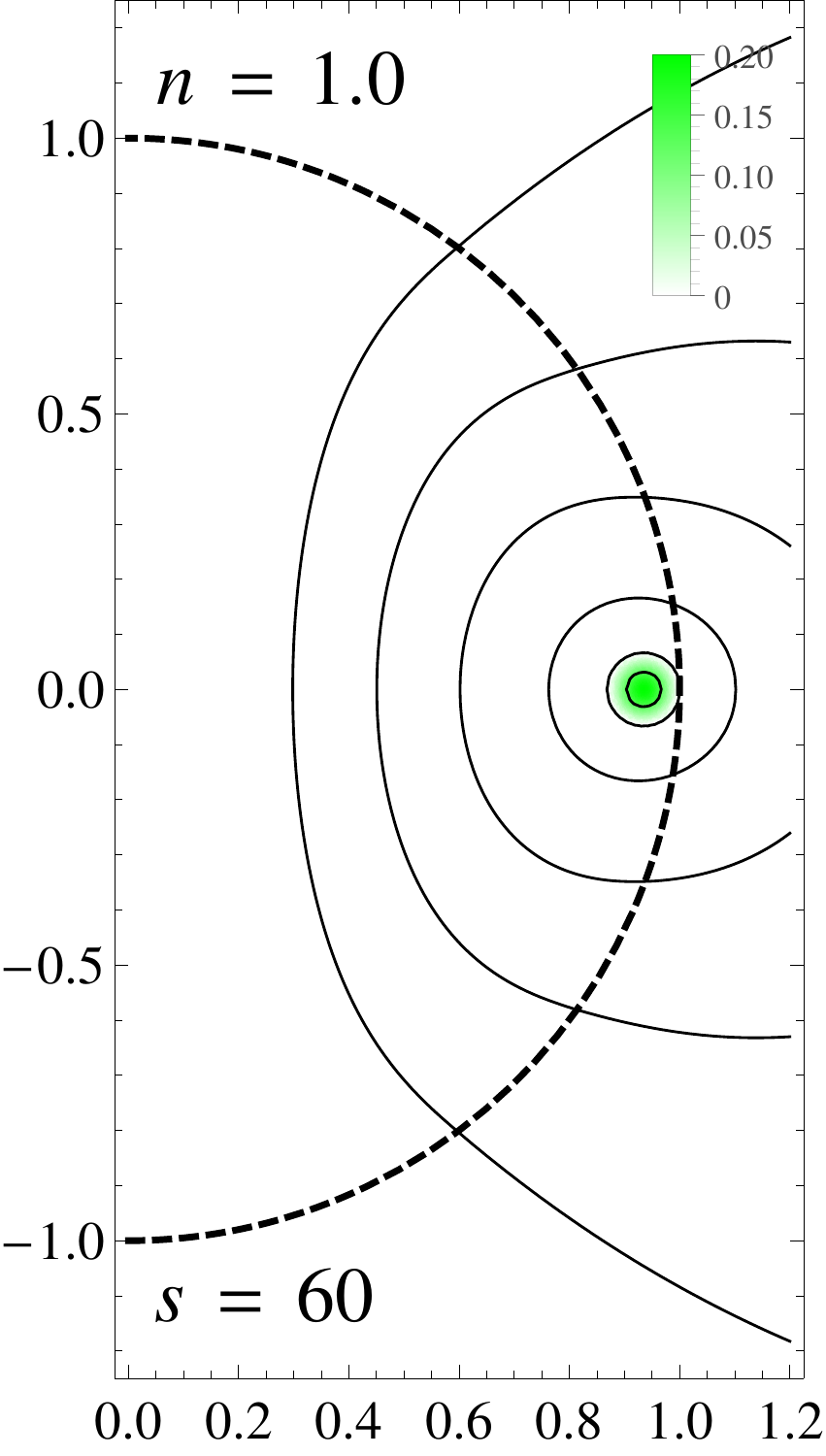}\\
\includegraphics[width=5.5cm]{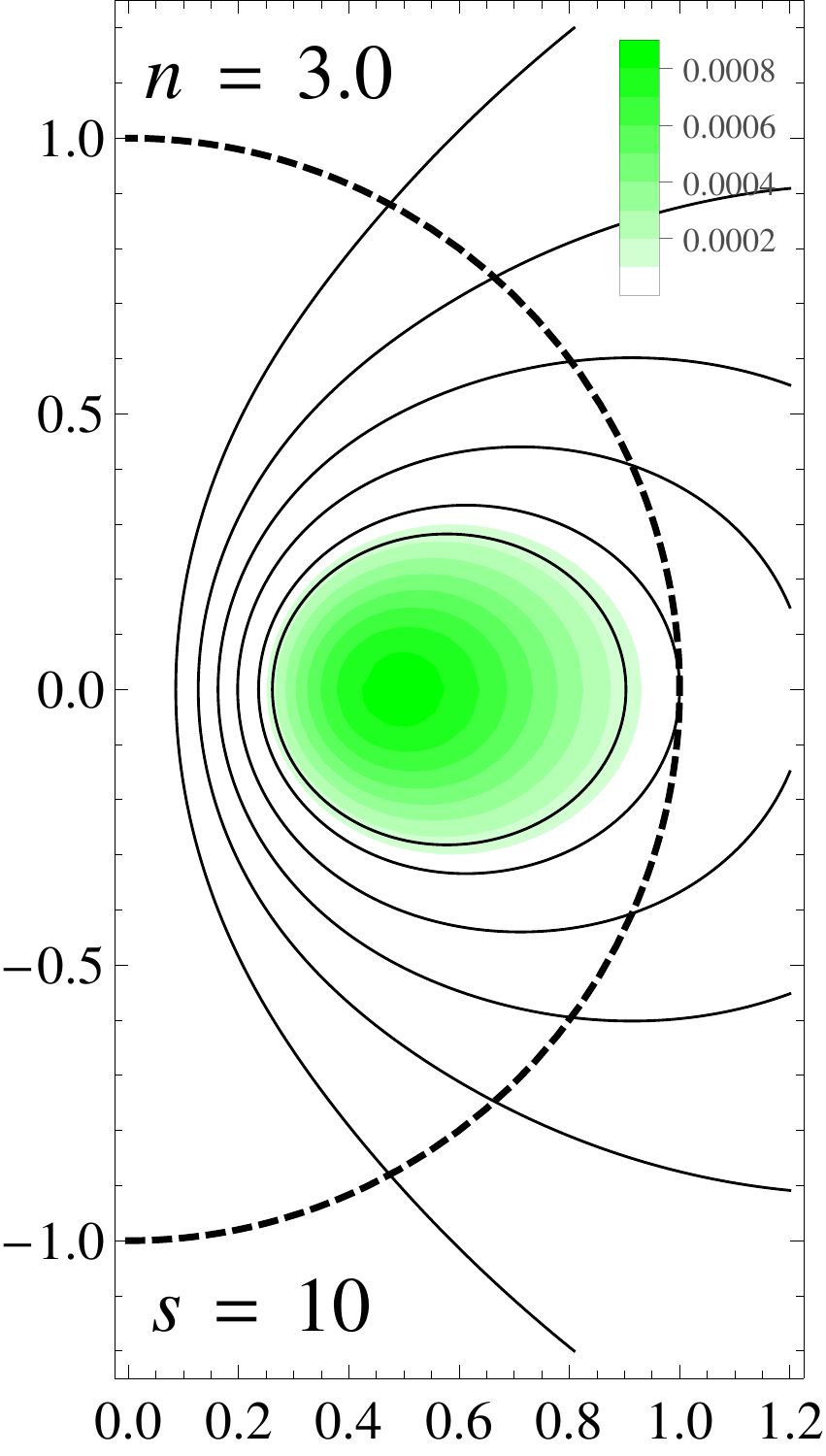}\hspace{1cm}\includegraphics[width=5.5cm]{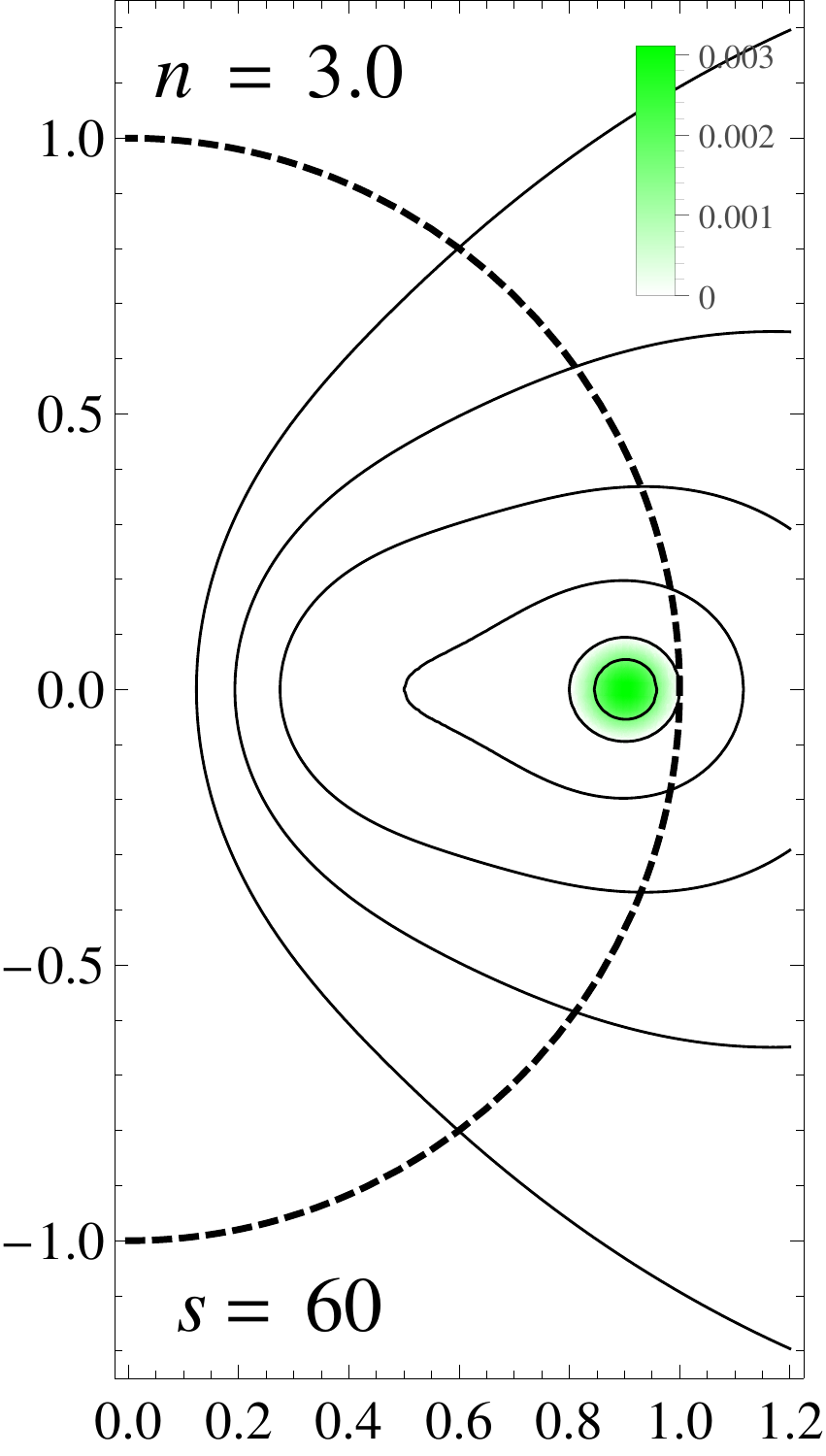}
\caption{Magnetic field of barotropic stars with polytropic equations of state, for $\chi'(\alpha) = 1$ and $\beta(\alpha)$ given in Eq. \eqref{twisted}. The dashed line is the surface, while solid lines are poloidal field lines corresponding to $0.2\alpha_\text{surf}$, $0.4\alpha_\text{surf}$, $0.6\alpha_\text{surf}$, $0.8\alpha_\text{surf}$, $\alpha_\text{surf}$ and $1.1\alpha_\text{surf}$, where $\alpha_\text{surf} = \alpha(R,\pi/2)$. The color map represents for $\beta(r,\theta) = r\sin\theta B_\phi(r,\theta)$. The axes are measured in units of the stellar radius. For these cases, we took $N_r = 400$ and $N_\theta = 500$.}\label{magpolyt2}
\end{center}
\end{figure*}

As illustrated in Fig. \ref{magpolyt1} for the particular case of $n=2$ polytropes, for small and moderate values of $s$, we found that the strength of the poloidal component is almost one order of magnitude higher than the toroidal one, while the poloidal field itself is around 10 times stronger near the center of the star compared to the field at the surface. For the latter range of $s$, the configurations are mostly dipolar, with almost negligible relative amplitude of higher multipoles.  A higher value of $s$ produces a configuration with a more substantial contribution of higher multipoles, although the dipole still dominates. As expected, increasing $s$ also increases the relative strength of the toroidal field with respect to the poloidal one, so, eventually, the amplitude of the toroidal field becomes comparable to the poloidal one. However, as Fig. \ref{magpolyt2} illustrates, the volume where the toroidal component is present shrinks in all cases, forming a thin ring of nearly circular cross section tangent to the surface of the star, a result already reported in the literature \citep{lander9, fujisawa12, gourgouliatos13, pili14}. Mathematically, this shrinkage can be understood as follows: in the GS equation, the term $r^2\sin^2\theta\rho\chi'$ is an explicit function of the position but not of $\alpha$ when $\chi'=\,$constant. In that case, and considering that $\beta\beta'$ increases when choosing a larger $s$, the term $\Delta^*\alpha$ must become more negative because $r^2\sin^2\theta\rho\chi'$ remains the same regardless of $s$. In turn, the term $\Delta^*\alpha$ is similar to the Laplacian of $\alpha$ (recall Eq. \eqref{GSop}), that is, it depends on the curvature of $\alpha$, so the larger $|\Delta^*\alpha|$ is, the more compact the poloidal lines are and the steeper $\alpha$ the maximum in $\alpha$ will be. The circular shape of the cross section occurs because, in the limit of $\sqrt{\text{cross section area}}\ll\text{radius of the ring}$, the toroidal current flowing along the ring is locally a straight line around which, since there is no matter to deform them, poloidal lines must form circular rings. This result is in agreement with the numerical results of \cite{lander9}, \cite{fujisawa12}, \cite{gourgouliatos13}, \cite{pili14} and \cite{bucciantini15}, although \cite{ciolfi9} find a different behavior.

\subsection{Fraction of the energy in the toroidal component}

\begin{figure}
\begin{center}
\includegraphics[width=8.1cm]{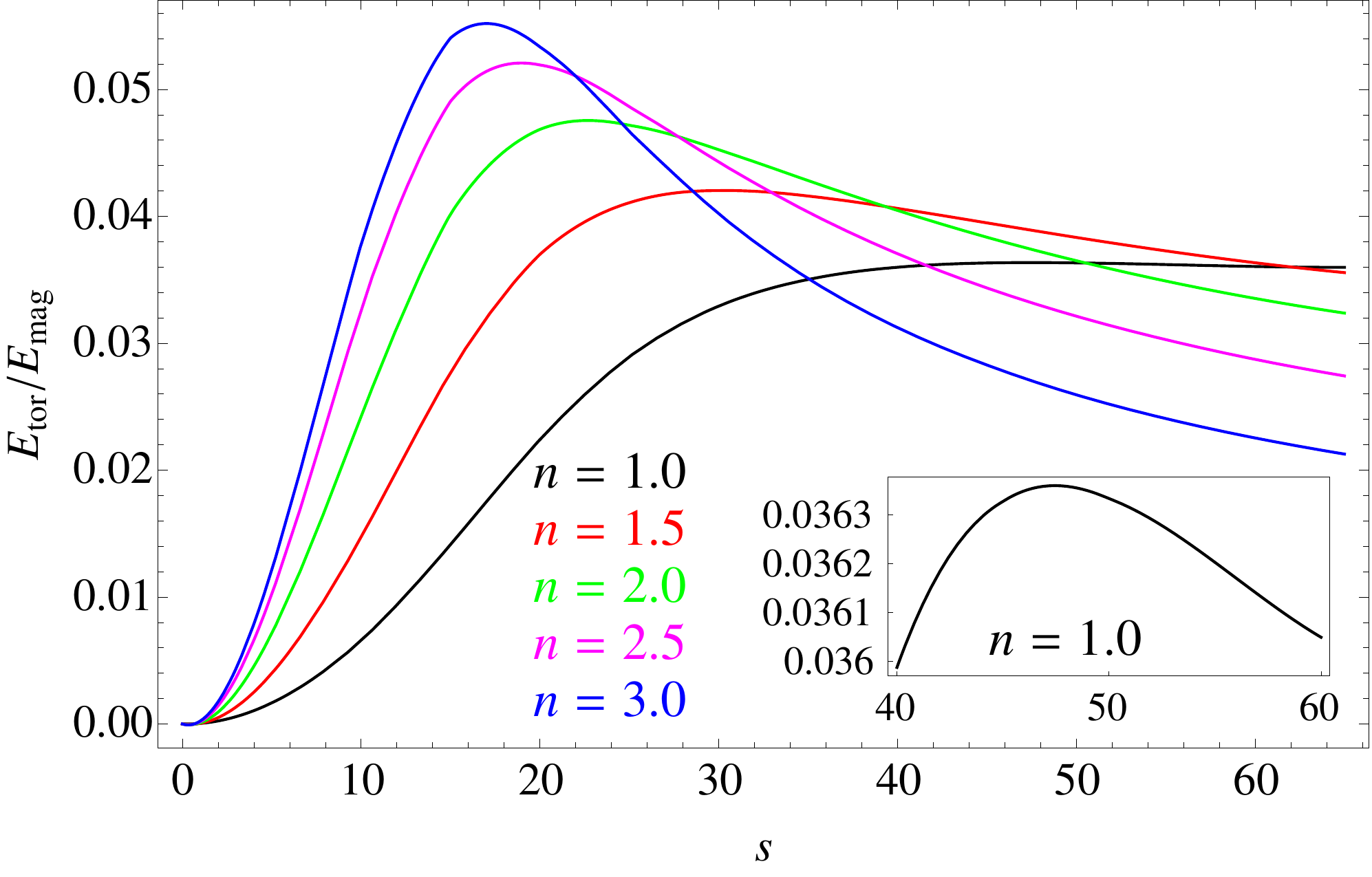}
\caption{Fraction of energy stored in the toroidal component of the magnetic field as function of the parameter $s$, for different polytropic equilibria explored in this work. For these cases, we took $N_r = 400$ and $N_\theta = 500$. Inset: blow-up of the region around the maximum for $n = 1$.}\label{rtor}
\end{center}
\end{figure}

Table \ref{energies} shows the fraction of toroidal energy for the magnetic polytropes described before. In all cases, this fraction is only a few \%{} of the total energy. Moreover, Fig. \ref{rtor} suggests that $E_\text{tor}/E_\text{mag}$ is bounded by a maximum value when plotted as a function of the parameter $s$, regardless of the density profile (see also \citealt{ciolfi9}, \citealt{lander9}, \citealt{fujisawa12}, \citealt{gourgouliatos13}, and \citealt{pili14}). Table \ref{smaxs} shows the values of $s_\text{max}$ at which the maxima occur.

\begin{table}
\begin{center}
\begin{tabular}{|c||cccccc|}
\hline
	&	$s=10$	&	20	&	30	&	40	&	50	&	60\\
\hline
$n=1.0$	&	0.657	&	2.24	&	3.29	&	3.60	&	3.63	&	3.61\\
$n=1.5$	&	1.47	&	3.70	&	4.20	&	4.06	&	3.83	&	3.63\\
$n=2.0$	&	2.43	&	4.69	&	4.52	&	4.05	&	3.65	&	3.35\\
$n=2.5$	&	3.27	&	5.19	&	4.43	&	3.71	&	3.21	&	2.87\\
$n=3.0$	&	3.78	&	5.34	&	4.02	&	3.12	&	2.59	&	2.25\\
\hline
\end{tabular}
\caption{Fraction of energy (in \%{}) stored in the toroidal component of magnetic equilibria with polytropic equation of state studied in this work. For these cases, we took $N_r = 400$ and $N_\theta = 500$.}\label{energies}
\end{center}
\end{table}

\begin{table}
\begin{center}
\begin{tabular}{|c|ccccc|}
\hline
	&	$n=1.0$	&	1.5	&	2.0	&	2.5	&	3.0\\
\hline
\hline
$(E_\text{tor}/E_\text{mag})_\text{max}$	&	3.64	&	4.20	&	4.76	&	5.21	&	5.52\\
$s_\text{max}$	&	47.5	&	29.9	&	22.7	&	19.0	&	17.0\\
\hline
\end{tabular}
\caption{Maximum fraction (in \%{}) of magnetic energy stored in the toroidal component for different polytropes, as functions of $s$. $s_\text{max}$ is the value of $s$ at which these maxima occur. For these cases, we took $N_r = 400$ and $N_\theta = 500$.}\label{smaxs}
\end{center}
\end{table}

The existence of a maximum value for $E_\text{tor}/E_\text{pol}$ can be understood in terms of two competing effects: for low values of $s$, the poloidal field is hardly affected by the weak toroidal field. Thus, the volume containing the latter does not change much with $s$, $E_\text{pol}\approx\,\text{const.}$, and $E_\text{tor}\propto \beta^2 \propto s^2$, so
\begin{equation}
\frac{E_\text{tor}}{E_\text{mag}} = \frac{E_\text{tor}}{E_\text{pol} + E_\text{tor}} \approx \frac{E_\text{tor}}{E_\text{pol}} \propto s^2,
\end{equation}
i.e., the toroidal ratio increases quadratically as a function of $s$. As $s$ increases more, so does the toroidal field, and the dependence with $s$ becomes less trivial. Since the region where this component lies shrinks, the volume integral defining $E_\text{tor}$ will stop increasing at some $s_\text{max}$ and the toroidal fraction will start to decrease. In other words, the interplay between increasing $|\mathbf B_\text{tor}|^2$ while shrinking the volume where it is present produces an upper bound for $E_\text{tor}/E_\text{mag}$. These results are qualitatively still true for other equations of state \citep{fujisawa12} and when relativistic effects are included \citep{ciolfi9, pili14}.

\subsection{On the validity of ``weak'' magnetic field}\label{weak}

All our calculations are based on the assumption of a \textit{weak} magnetic field, in the sense that the magnetic forces do not exert a significant effect on the stellar structure, i.e., to first order, those forces do not deform the primarily spherical shape of the star. Making use of this assumption, we took the simpler approach of obtaining magnetic equilibria by solving the GS equation only, assuming a fixed spherical density profile coming from a non-magnetic background star, instead of solving self-consistently the whole system of equations (Euler equation + Poisson equation + Maxwell equations) provided an equation of state $P = P(\rho)$, i.e., obtaining not only $\alpha(r,\theta)$ but also the fluid quantities, as functions of position. It is natural to wonder about the accuracy of the simpler approach used throughout this work compared to the self-consistent scheme. The latter method was already used by Lander \& Jones (2009, 2012) (L\&{}J) in the same astrophysical context. Table \ref{withLander} exhibits a comparison of the fraction of toroidal energy for equilibria obtained by L\&{}J versus our work, for two different polytropic indices. With these values, we can estimate the expected discrepancy in neglecting the effect of the magnetic field on the fluid as $E_\text{mag}/E_\text{grav} \sim 10^{-3}$, which is in acceptable agreement with the relative error shown in Table \ref{withLander}. Since the magnetic field strength needed to generate a minimum distortion on the fluid is at least one order of magnitude above the fields for the objects considered here, then the approximation discussed should hold for the whole relevant range of magnetic fields. This has the implication of enormously simplifying the process of finding barotropic equilibria.

\begin{table}
\begin{center}
\begin{tabular}{|c|cccc|}
\hline
	&	$1.1\,s$	&	$k$	&	L\&{}J	&	this work\\
\hline
\hline
\multirow{5}{*}{$n=1$}	&	10	&	1.574	&	0.586	&	0.596\\
	&	20	&	1.39	&	2.07	&	2.05\\
	&	30	&	1.254	&	3.18	&	3.16\\
	&	40	&	1.189	&	3.56	&	3.57\\
	&	50	&	1.157	&	3.55	&	3.64\\
\hline
\hline
\multirow{3}{*}{$n=3$}	&	10	&	3.24	&	3.87	&	3.89\\
	&	20	&	2.55	&	5.30	&	5.32\\
	&	25	&	2.46	&	4.62	&	4.69\\
\hline
\end{tabular}
\caption{Fraction of toroidal energy $E_\text{tor}/E_\text{mag}$ (in \%{}) obtained in this work vs. those obtained by Lander \& Jones (2009, 2012). The models calculated by L\&J assume a mass $M = 1.4 M_\odot$ and radius $R = 10\,\text{km}$, with a resulting average magnetic field $4.5\times 10^{16}\,$G for $n=1$ and $1.0\times 10^{16}\,$G for $n = 3$. In our simulations, we took $N_r = 400$ and $N_\theta = 500$.}\label{withLander}
\end{center}
\end{table}

\section{Asymptotic, analytic solutions}

Numerical instabilities arise when increasing $s$ beyond a value $\sim 65$ (for fixed $\chi' = 1$) and our code has convergence problems. This occurs because more points on the grid are needed to resolve 
the toroidal volume, which becomes smaller. In this section we overcome this difficulty by introducing an asymptotic model for large $s$, which allows us to study the global properties of the equilibria in the limit $s\longrightarrow \infty$.  

\begin{SCfigure}
\centering
\boxed{\includegraphics[width=0.15\textwidth]{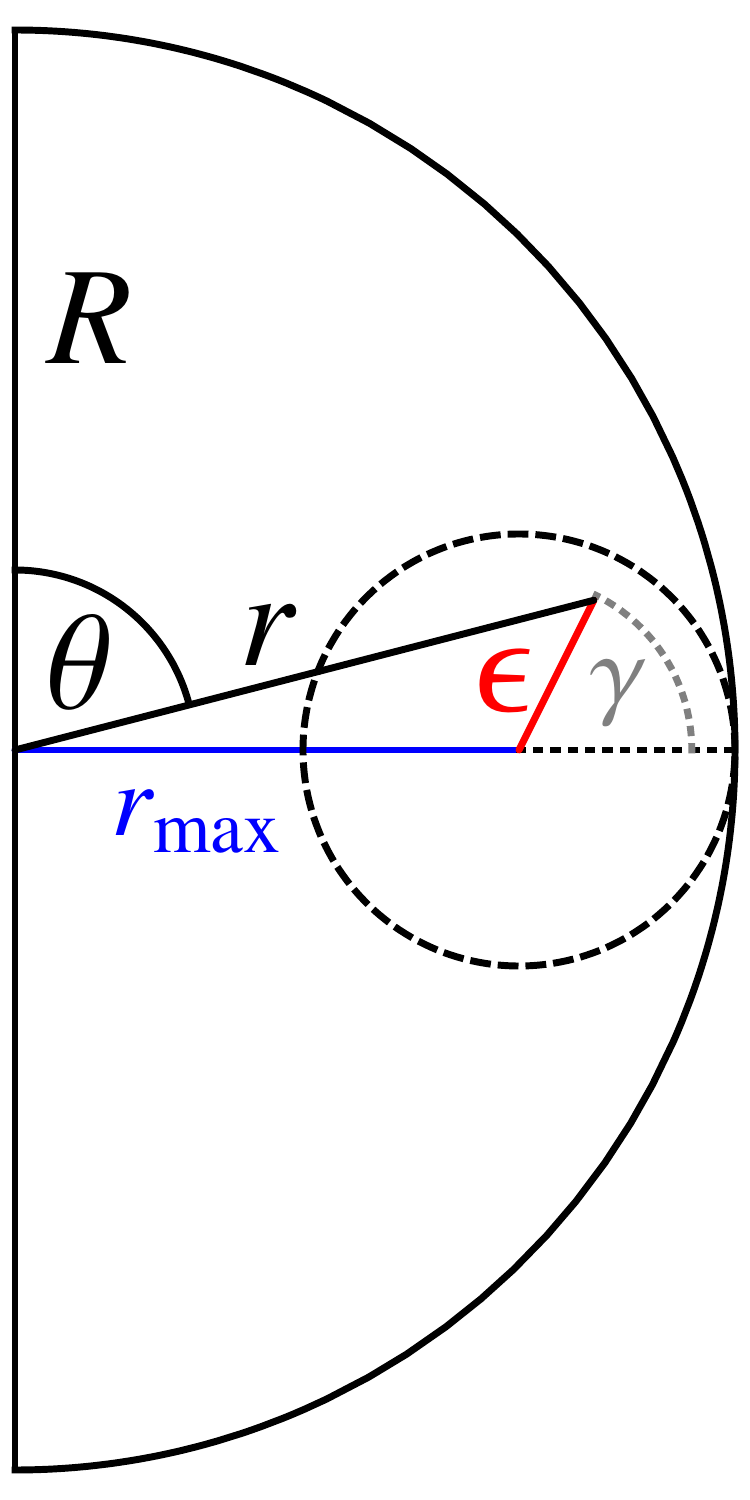}}
\caption{New coordinate system introduced to describe the volume where the toroidal field lies when increasing $s$ beyond $s_\text{max}$ (the cross section of the toroid is exaggerated for visualization).}\label{newcoords}
\end{SCfigure}

In order to motivate the model, we note that the term $r^2\sin^2\theta\rho\chi'$ in the GS equation approaches zero as one approaches the stellar surface, because it is proportional to $\rho$, which is expected to decrease monotonically until it vanishes at the surface. Thus, it is also expected that, as the volume occupied by the toroidal field shrinks and gets close to the surface, this term becomes significantly smaller in magnitude than the other two inside the volume, so that $\Delta^*\alpha$ and $\beta\beta'$ roughly cancel each other. In fact, from our simulations we confirmed that
\begin{equation}
\frac{r^2\sin^2\theta \rho}{|\Delta^*\alpha|} \sim \frac{r^2\sin^2\theta \rho}{\beta\beta'} \lesssim 10^{-3}
\end{equation}
for $s \gtrsim 60$ (again, we normalize $\chi' = 1$). In this way, one can expect that the relation
\begin{equation}\label{approxGS}
\Delta^*\alpha + \beta\beta' \approx 0
\end{equation}
will become more and more accurate inside the toroidal volume when increasing $s$. This is the central idea of the field model we are introducing: for large $s$, we assume that Eq. \eqref{approxGS} is valid, and solve for $\alpha$ under that approximation. Of course, when $\alpha$ is determined, the problem is solved and then one can obtain expressions for the toroidal flux, energy, and so on. The problem is even simpler if we recall that the volume where the toroidal field lies not only shrinks but also becomes a toroid regardless of the density profile. Because of this, we introduce a new coordinate system $(\epsilon, \gamma)$ on the meridional half plane of the star, see Fig. \ref{newcoords}. From our simulations, it turns out that the smaller the volume, the more independent $\alpha$ is with respect to the angle $\gamma$, so that for a sufficiently large $s$, we can assume that $\alpha$ does not depend on $\gamma$, but only on $\epsilon$, therefore the GS operator reads (recall Eq. \eqref{GSop})
\begin{equation}
\Delta^*\alpha  \approx \boldsymbol\nabla^2 \alpha  = \frac{1}{\epsilon}\frac{d}{d\epsilon}\left(\epsilon\,\frac{d\alpha}{d\epsilon} \right),
\end{equation}
where we have used that $r\sin\theta \approx\,\text{constant}\, (\approx r_\text{max})$. Using the function $\beta(\alpha)$ of Equation \eqref{twisted}, Equation \eqref{approxGS} inside the toroidal volume can be written as
\begin{equation}\label{quasiLE}
\frac{1}{u}\frac{d}{du}\left(u\frac{d\eta}{du}\right) = - \eta^{2\lambda - 1},
\end{equation}
where we have defined the dimensionless function $\eta$ so that $\alpha - \alpha_\text{surf} = (\alpha_\text{max} - \alpha_\text{surf})\eta$, with $\alpha_\text{max}$ being the maximum value of $\alpha$ (reached at $r = r_\text{max}, \theta = \pi/2$, or $\epsilon = 0$), and the dimensionless radius
\begin{equation}\label{defofu}
u \equiv \lambda^{1/2}(\alpha_\text{max} - \alpha_\text{surf})^{\lambda-1}\,s\epsilon.
\end{equation}
Eq. \eqref{quasiLE} is a second-order ordinary differential equation for $\eta = \eta(u)$, hence two boundary conditions are needed to solve it. Firstly, and by definition, it is clear that $\eta(0) = 1$. Also, $\eta'(0) = 0$, where the prime stands for the derivative with respect to the argument. This is because $\alpha$ (and thus $\eta$) has a smooth maximum at  $\epsilon = 0$. Eq. \eqref{quasiLE} must be integrated until the boundary of the toroidal volume, corresponding to $\alpha = \alpha_\text{surf}$, or  $\eta(u_\text{surf}) = 0$, where $u_\text{surf}$ is the first root of $\eta(u)$, a number to be determined. For $u > u_\text{surf}$ there is no toroidal field, so this equation (and its solution) is valid for $u \leq u_\text{surf}$ only.  Eq. \eqref{quasiLE} accepts analytical solutions for a few exponents $\lambda$. For $\lambda = 1/2$, the unique solution  satisfying the boundary conditions is $\eta(u) = 1- u^2/4$, with $u_\text{surf} = 2$. For $\lambda = 1$. the equation becomes a Bessel equation of order zero, with unique solution $\eta(u) = J_0(u)$, $u_\text{surf} \approx 2.4048$ being the first root. For all other values of $\lambda$, the equation is nonlinear, with no obvious analytical solution, so we wrote a fourth-order Runge-Kutta code to solve it numerically. Of course, $u_\text{surf}$ is a pure number, depending on the value of $\lambda$ \textit{and on nothing else}.

\subsection{Toroidal quantities}

Under the above assumptions, formul\ae{} for the toroidal flux and energy are readily obtained. By definition, the toroidal flux through a meridional plane is 
\begin{align}\nonumber\label{analflux}
\Phi_\text{tor}  &\approx \int_0^{\epsilon_\text{surf}}\frac{\beta(\epsilon)}{r_\text{max}}\,2\pi\,\epsilon\,d\epsilon\\ 
&= \frac{2\pi \,q_{\lambda}}{\lambda r_\text{max}s}(\alpha_\text{max} - \alpha_\text{surf})^{2-\lambda},
\end{align}
where we defined 
\begin{equation}\label{Iq}
q_\nu \equiv \int_0^{u_\text{surf}}u\eta^\nu\,du,
\end{equation}
a pure number. The toroidal energy can be computed as
\begin{align}\nonumber\label{analenergy}
E_\text{tor} &\approx \int_0^{\epsilon_\text{surf}}\frac{\beta^2(\epsilon)}{8\pi r_\text{max}^2}\,4\pi^2 r_\text{max}\,\epsilon\,d\epsilon\\
 &= \frac{\pi \,q_{2\lambda}}{2\lambda r_\text{max}}(\alpha_\text{max} - \alpha_\text{surf})^2.
\end{align}

In principle, we are interested in obtaining analytical expressions as functions of the parameter $s$. In Eqs. \eqref{analflux} and \eqref{analenergy}, $\alpha_\text{max}$, $\alpha_\text{surf}$, and $r_\text{max}$ (which to first approximation equals the radius $R$) depend implicitly on $s$, so it is desirable to obtain these dependences. This can be done as follows. First, let us consider the hypothetical case in which the toroidal field is effectively concentrated in an infinitesimally thin loop at the surface along the equator of the star. This would be the ``$s\longrightarrow \infty$'' case. This limit of an infinitesimally thin loop is analogous to the problem of a thin, circular current loop. Outside this loop, $\alpha$ is given by the solution of the GS equation for a purely poloidal field, namely,
\begin{equation}
\alpha = \begin{cases}
\displaystyle\sum_{\ell = 1}^\infty\left[a_\ell \left(\frac{r}{R}\right)^{\ell + 1} - f(r)\delta_{\ell 1}\right]\sin\theta P_\ell^1(\cos\theta) & r < R\\
\displaystyle\sum_{\ell = 1}^\infty b_\ell \left(\frac{R}{r}\right)^\ell\sin\theta P_\ell^1(\cos\theta) & r > R,\label{generalsolution}
\end{cases}
\end{equation}
where $\delta_{\ell m}$ is the Kronecker delta and
\begin{equation}
f(r) = \frac{k}{3r}\int_{0}^r\xi\rho(\xi)(\xi^3 - r^3)\,d\xi,
\end{equation}
where $\chi' = k$ (a constant) regulates the amplitude of the magnetic field \citep{gourgouliatos13}. The coefficients $a_\ell$ and $b_\ell$ must be fixed using boundary conditions. Continuity of $\alpha$ at $r =R$ implies
\begin{equation}
a_\ell - f(R)\delta_{\ell 1} = b_\ell.\label{aellbell}
\end{equation}
This also ensures continuity of the radial component of the magnetic field. This time, however, the tangential component of the magnetic field is not continuous at the boundary $r=R$, but it has a discontinuity due to the presence of a thin loop carrying a current $I$, with current density
\begin{equation}
\mathbf J = \frac{I}{R}\,\delta(\cos\theta)\,\delta(r - R)\mathbf e_\phi.\label{Jthinloop}
\end{equation} 
Here $\delta(x)$ is the Dirac delta function. This current comes from a singular toroidal field, which produces an azimuthal current density
\begin{equation}
\frac{4\pi}{c} J_\phi = \left.(\boldsymbol\nabla\times\mathbf B)\right|_\phi = - \frac{\Delta^*\alpha}{r\sin\theta} = \frac{\beta\beta'}{r\sin\theta},\label{Jthinloop2}
\end{equation} 
where in the last equality we have used Eq. \eqref{approxGS}. The second boundary condition to impose is
\begin{equation}
\left.\frac{\partial \alpha}{\partial r}\right|_{R^+} - \left.\frac{\partial \alpha}{\partial r}\right|_{R^-} = - \frac{4\pi I}{c}\,\sin\theta\,\delta(\cos\theta).\label{discradialder}
\end{equation}
By direct replacement of the respective expressions for $\partial\alpha/\partial r$ for $r < R$ and for $r> R$, one can extract the coefficients by standard mathematical tools in order to finally get
\begin{equation}
a_\ell = \frac{f(R) + Rf'(R)}{3}\delta_{\ell 1} + \frac{2\pi I R}{c}\frac{P_\ell^1(0)}{\ell(\ell + 1)},\label{aelllimit}
\end{equation}
from which $b_\ell$ can be obtained from Eq. \eqref{aellbell}. Thus, if $I$ is known, the problem in this case is completely determined.

The motivation behind pointing out this hypothetical case is what follows. The general solution in Eq. \eqref{generalsolution} is of course also valid outside the toroidal region when the latter is a thin (but not infinitesimally thin) loop. In that case, however, the coefficients are not readily obtained as we would have to impose boundary conditions in a cumbersome geometry, assuming that the solution to the GS equation inside the toroidal volume is precisely that satisfying Eq. \eqref{quasiLE}. Nevertheless, as the parameter $s$ increases, the coefficients should eventually converge to those in Eq. \eqref{aelllimit}. Therefore, in the limit of a  thin toroidal volume with finite cross section, it is expected that the solution to the GS equation will be that in Eq. \eqref{generalsolution} with coefficients $a_\ell$ and $b_\ell$ given \emph{approximately} by those in \eqref{aelllimit}, where $I$ is the current along the thin toroidal region. If this is the case, the current $I$ flowing can be easily obtained using the asymptotic model developed in this section. By definition, and recalling Eq. \eqref{Jthinloop2}, we get
\begin{align}\nonumber
I &\approx \frac{c}{4\pi}\int_0^{\epsilon_\text{surf}}\frac{\lambda s^2 (\alpha_\text{max} - \alpha_\text{surf})^{2\lambda - 1}}{r_\text{max}}\eta^{2\lambda - 1}\,2\pi\,\epsilon\,d\epsilon\\
&= \frac{c \,q_{2\lambda-1}}{2R}(\alpha_\text{max} - \alpha_\text{surf}),\label{Ilimit}
\end{align}
where we have replaced $r_\text{max}\approx R$. Eq. \eqref{Ilimit} implies that in this approximation, $I$, and hence the coefficients, are implicit functions of $s$. The asymptotic solution of the GS equation for large $s$, outside the toroidal volume, is that given in Eq. \eqref{generalsolution}, where the coefficients are approximately
\begin{equation}
a_\ell \approx \frac{f(R) + Rf'(R)}{3}\delta_{\ell 1} + q_{2\lambda -1}(\alpha_\text{max} - \alpha_\text{surf})\frac{\pi P_\ell^1(0)}{\ell(\ell + 1)}
\end{equation}
and
\begin{equation}
b_\ell \approx \frac{Rf'(R) - 2f(R)}{3}\delta_{\ell 1} + q_{2\lambda -1}(\alpha_\text{max} - \alpha_\text{surf})\frac{\pi P_\ell^1(0)}{\ell(\ell + 1)}.
\end{equation}
So, with this procedure, we can write an approximate solution $\alpha$ for the GS equation in terms of its maximum value $\alpha_\text{max}$ and its value $\alpha_\text{surf}$ at the surface between the region with and without toroidal field. In addition, $\alpha$ and its derivatives must be continuous across the latter surface, i.e.,
\begin{equation}
\alpha_\text{surf} = \alpha(R-\epsilon_\text{surf},\pi/2) \label{cond11}
\end{equation}
 and
\begin{equation}
\frac{\alpha_\text{max} - \alpha_\text{surf}}{\epsilon_\text{surf}}q_{2\lambda-1} = \frac{\partial\alpha}{\partial r}(R-\epsilon_\text{surf},\pi/2),\label{cond12}
\end{equation}
respectively. Introducing the explicit form of the coefficients $a_\ell$ in the expansion for $\alpha$ (Equation \eqref{generalsolution}), Equations \eqref{cond11} and \eqref{cond12} read

\begin{widetext}
\begin{equation}
\alpha_\text{surf} = k\rho_c R^4\left[f(1-x) - \frac{f(1) + f'(1)}{3}(1-x)^2\right] + \pi q_{2\lambda -1}(\alpha_\text{max} - \alpha_\text{surf})F(1-x)\label{cond11final}
\end{equation}
and
\begin{equation}
\frac{\alpha_\text{max} - \alpha_\text{surf}}{x} = \frac{k\rho_c R^4}{q_{2\lambda-1}} \left[f'(1-x) - \frac{2(f(1) + f'(1))}{3}(1-x)\right] + \pi (\alpha_\text{max} - \alpha_\text{surf})F'(1-x),\label{cond12final}
\end{equation}
\end{widetext}
respectively, where we have redefined $f(r) \equiv k\rho_c R^4\, \tilde f(r/R)$ and then dropped the tilde. Also, we have introduced the parameter $x = \epsilon_\text{surf}/R$ and the function 
\begin{align}\nonumber
F(z) &\equiv \sum_{\ell = 1}^\infty\frac{[P_\ell^1(0)]^2}{\ell(\ell + 1)}z^{\ell + 1} = z^2\sum_{\nu = 0}^\infty \frac{2\nu + 1}{2\nu + 2}\left[P_{2\nu}(0)\right]^2 z^{2\nu}\\
 &= \frac{z^2}{8}\left[ 4\, _2F_1\left(\frac{1}{2},\frac{1}{2},2,z^2\right) + z^2\, _2F_1\left(\frac{3}{2},\frac{3}{2},3,z^2\right)\right].
\end{align}
Here $_2 F_1(a,b,c,z)$ is the Gauss hypergeometric function, so that $F(z)$ converges for $|z|<1$, while $F'(1-x)$ means derivative $F'(z)$ evaluated at $z = 1-x$. The interesting point with the previous formul\ae{} is that Eqs. \eqref{cond11final} and \eqref{cond12final} allow to write both $\alpha_\text{max}$ and $\alpha_\text{surf}$ in terms of the parameter $x = \epsilon_\text{surf}/R$, which is an implicit function of $s$. Once one solves for $\alpha_\text{max}(x)$ and $\alpha_\text{surf}(x)$, along with Eq. \eqref{defofu} evaluated at $\epsilon_\text{surf}$,
\begin{equation}
u_\text{surf} = \lambda^{1/2}(\alpha_\text{max} - \alpha_\text{surf})^{\lambda-1}s \epsilon_\text{surf},\label{u_s}
\end{equation} 
where $u_\text{surf}$ is a pure number, one gets the equations needed to write $\Phi_\text{tor}$ and $E_\text{tor}$ as a function of $s$. Since by construction the poloidal flux equals $2\pi\alpha$, we can define the quantity $\Phi_\text{pol} = 2\pi\alpha_\text{max}$, which is the maximum poloidal flux reached by the configuration, and take the ratio $\Phi_\text{tor}/\Phi_\text{pol}$. Using \eqref{analflux}, we get
\begin{equation}
\frac{\Phi_\text{tor}}{\Phi_\text{pol}} = \frac{q_\lambda}{\lambda^{1/2}u_\text{surf}}\frac{x}{1-x}\frac{\alpha_\text{max}(x) - \alpha_\text{surf}(x)}{\alpha_\text{max}(x)},\label{rfluxfinal}
\end{equation}
which turns out to be an explicit function of $x= \epsilon_\text{surf}/R$. This equation can be combined with 
\begin{equation}
s = \frac{u_\text{surf}}{\lambda^{1/2}R\,x\,(\alpha_\text{max}(x) - \alpha_\text{surf}(x))^{\lambda-1}},\label{sfinal}
\end{equation}
(see Eq. \eqref{u_s}) which is also an explicit function of $x$, to obtain a parametric relation between $s$ and $\Phi_\text{tor}/\Phi_\text{pol}$. Eq. \eqref{rfluxfinal} allows to compare this model with the simulations.

\subsection{The model versus the simulations} 

\begin{figure}
\begin{center}
\includegraphics[width=6.9cm]{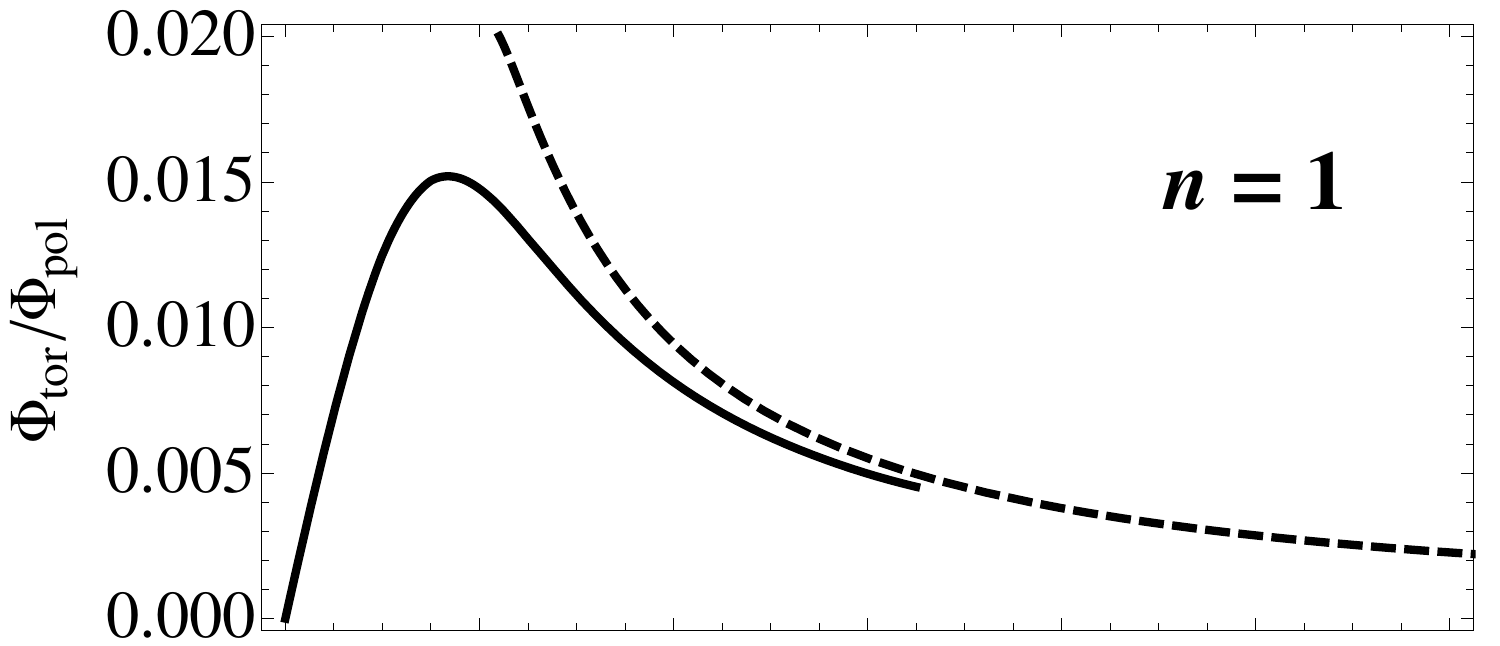}\\\vspace{0.12cm}
\hspace{0.13cm}\includegraphics[width=6.75cm]{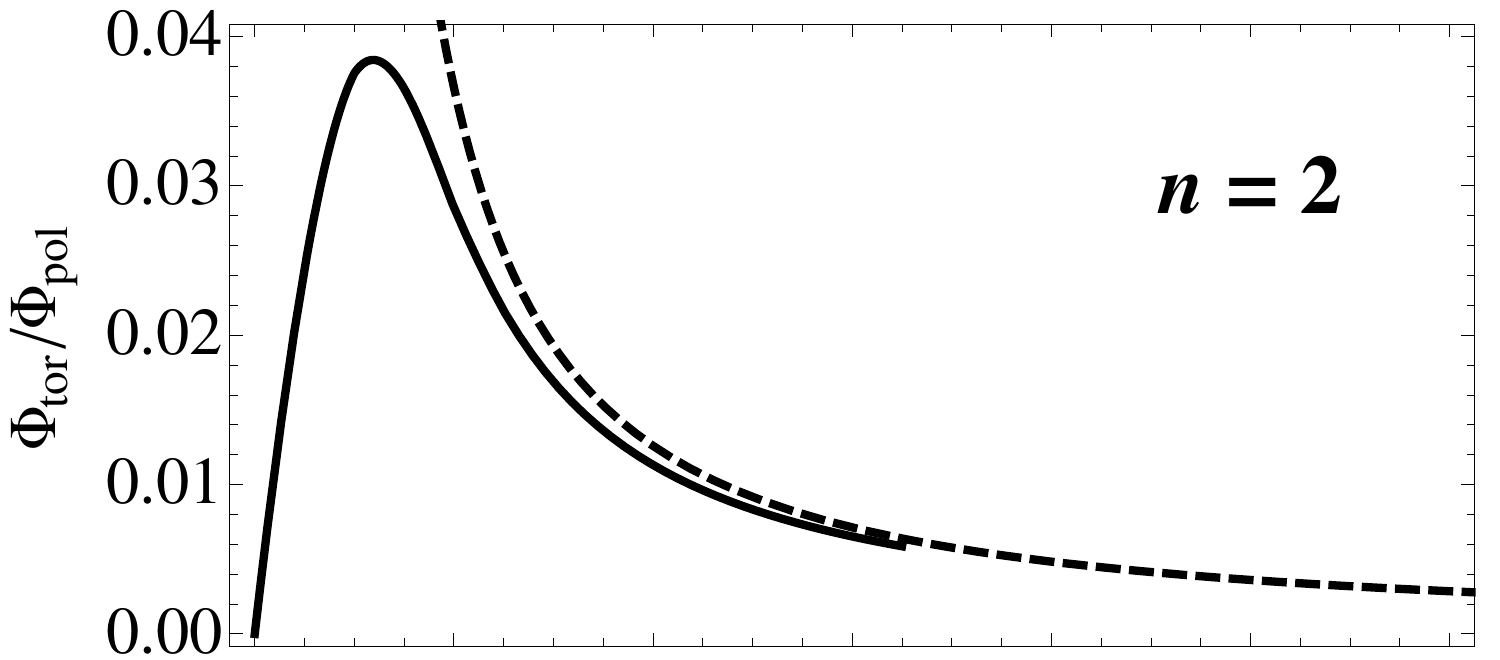}\\\vspace{0.12cm}
\hspace{0.13cm}\includegraphics[width=6.75cm]{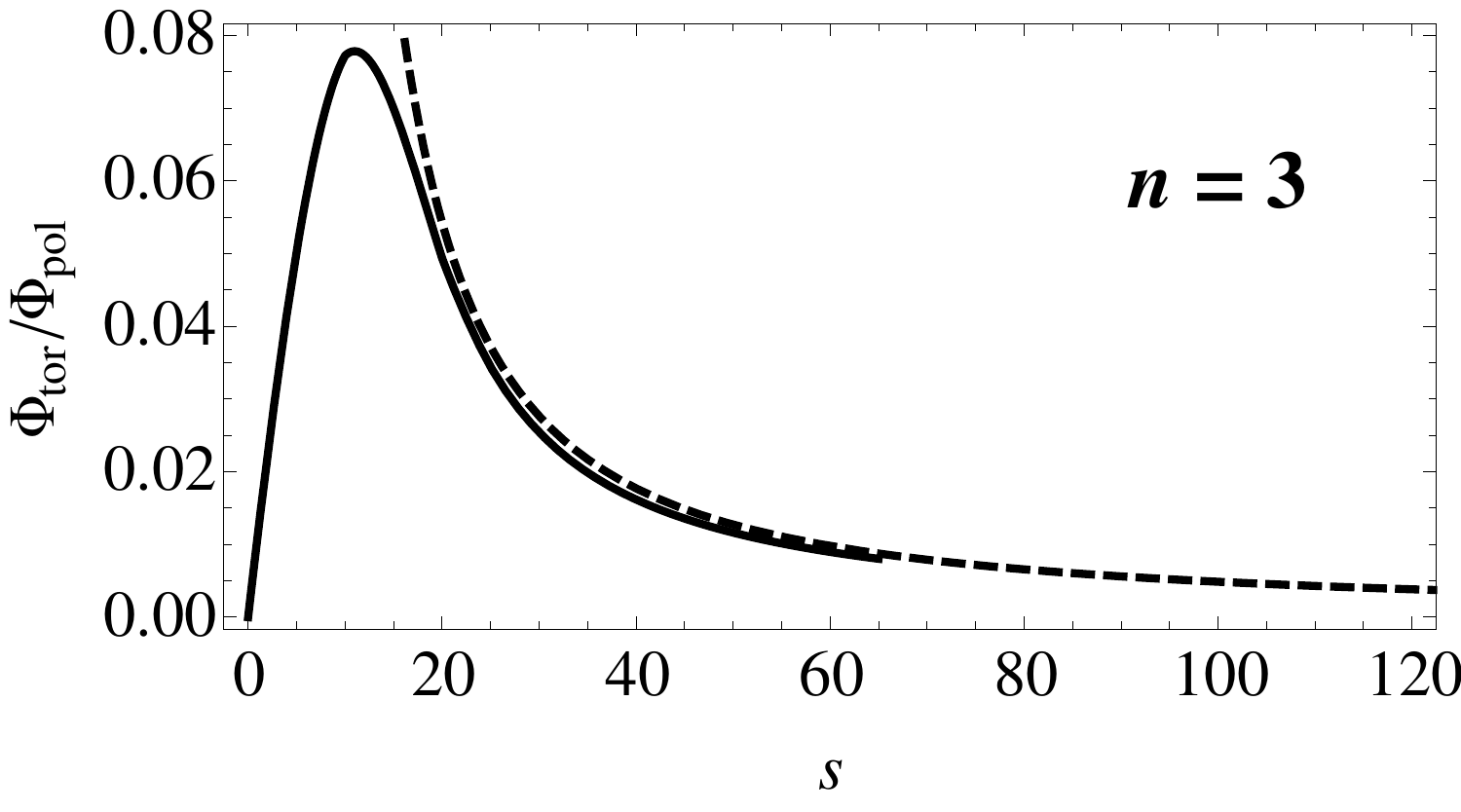}
\caption{Ratio of magnetic fluxes of the two components of the magnetic field, as a function of the parameter $s$, for different polytropic equations of state. The solid line are numerical simulations, while the dashed line is the evaluation of eqs. \eqref{rfluxfinal} and \eqref{sfinal} corresponding to the analytic model.}\label{ajustefinal}
\end{center}
\end{figure}

As discussed, the limit of having an infinitesimally thin toroidal volume is reached when $x\longrightarrow 0$, or $s\longrightarrow \infty$. Fig. \ref{ajustefinal} displays $\Phi_\text{tor}/\Phi_\text{pol}$ vs. $s$ obtained from eqs. \eqref{rfluxfinal} and \eqref{sfinal}, as well as from the numerical simulations. In all cases, as $s$ increases,  the asymptotic model approaches the numerically obtained curve, in good agreement with the approximations assumed within the model. 
Having tested the model with the simulations, the question whether $E_\text{tor}/E_\text{mag}$ is bounded over the full range of $s$ can be answered. In this case, there is no trivial expression for $E_\text{pol}$ as one should in principle integrate the asymptotic solution over all space outside the thin toroidal volume, and add the contribution inside the torus. This would demand non-trivial integration limits for the former integral, although the latter can be easily computed with the model of this section. In order to estimate the poloidal energy, we can use the known result for a circular loop of radius $r_\text{max}$ having  a circular cross section of radius $\epsilon_\text{surf}$, with $\epsilon_\text{surf} \ll r_\text{max}$ (e.g. \citealt{jackson75})
\begin{equation}
E_\text{pol} = \frac{1}{2}L I^2;\quad L \approx \frac{4\pi R}{c^2}\left[\ln\left(\frac{8 R}{\epsilon_\text{surf}}\right) - \frac{7}{4}\right],
\end{equation}
which diverges in the limit $\epsilon_\text{surf}/R\longrightarrow 0$. The toroidal energy, on the other hand, goes to zero as $x\longrightarrow 0$, so the fraction $E_\text{tor}/E_\text{pol}$ also vanishes in that limit. The conclusion is that $E_\text{tor}/E_\text{pol}$, \emph{as a function of $s$, does possess a maximum value, and then decreases monotonically to zero}. This would mean that \emph{all} the configurations with $\beta(\alpha)$ as given in Eq. \eqref{twisted} and $\chi'(\alpha) = \,\text{const.}$ are restricted to have a small ($\lesssim 10\%{}$) fraction of the magnetic energy stored in the toroidal component. This choice has been the most used in the literature, so if one were interested in obtaining higher fractions, one should explore other prescriptions for $\chi'(\alpha)$ (and perhaps for $\beta(\alpha)$ as well).

\section{Summary and conclusions}

We have developed a numerical code to model axially symmetric magnetic stars in ideal MHD equilibrium, allowing for both poloidal and toroidal component of the magnetic field, in a barotropic fluid. We presented several tests to check its accuracy, including comparisons with previous works in this area, finding a good agreement with them. From these comparisons, we showed that, for magnetic field strengths up to $\sim 10^{16}$ G, and assuming polytropic equations of state, the simpler approach of solving the resulting GS equation describing the equilibrium, and imposing a density profile derived from a non-magnetic equilibrium, gives essentially the same results as when obtaining equilibria with the self-consistent scheme of solving not only for the magnetic field, but also for the fluid quantities provided an equation of state describing the fluid. 

Using our code, we found a relatively wide range of barotropic equilibria whose magnetic fields combine an internal toroidal field and a poloidal field extending to the exterior of the star as well. Fixing $\chi'(\alpha) = \,\text{const.}$, numerical equilibria described by the piecewise function $\beta(\alpha)$ in Eq. \eqref{twisted} are poloidal-dominated, in the sense that the fraction of magnetic energy stored in the toroidal component of the magnetic field is $\lesssim 10\%{}$, even for configurations with comparable poloidal and toroidal magnetic field strength. Our numerical simulations also confirm previous results showing a maximum value of this fraction as a function of the parameter $s$ appearing in $\beta(\alpha)$. These properties do not depend significantly on the density profile assumed, which in this work varied between several very different polytropes.

Numerical simulations break down for large  values of $s$ ($\gtrsim 65$). In order to obtain results beyond this limitation, we developed an analytical model which reproduces the behavior of the numerical solutions for large toroidal field. This model was able to mimic the main global properties in this regime, finding that both the fractions of magnetic fluxes and energies do have a maximum when plotted against $s$, and decay to zero when $s$ goes to infinity. 

It is likely that functions providing a larger region of poloidal field lines closing inside the star give larger toroidal energy, as proposed by \cite{ciolfi13}. This is something we can readily explore as our code accepts arbitrary functions $\beta(\alpha)$ and $\chi'(\alpha)$. Also, having obtained barotropic equilibria allows to study their stability. This can be done either through time-dependent MHD simulations using the configurations we have obtained as initial condition, or by means of analyzing the change of the energy in the system when slightly perturbing the fluid inside the star, in an analogous treatment to that for non-barotropic, stably stratified stars followed by \cite{akgun13}. 

\section*{Acknowledgments}

This work was supported by CONICYT International Collaboration Grant DFG-06,
FONDECYT Regular Grants 1110213, 1110135, 1150411 and 1150718, the Basal Center for Astrophysics 
and Associated Technologies (CATA; PFB-06), and a CONICYT Master's Fellowship.
 We thank K. N. Gourgouliatos, S. K. Lander, K. Fujisawa, and P. Marchant
 for having provided part of their results, which yielded very stimulating discussions, 
and the anonymous referee for useful comments on the first draft of this paper.

\appendix

\section{Details on our numerical code}

We are interested in obtaining the function $\alpha(r,\theta)$ that solves the generalized GS equation, 
\begin{equation}\label{genGSeq}
\Delta^*\alpha = -F\left(r,\theta,\alpha(r,\theta)\right),\quad\quad F(r,\theta,\alpha(r,\theta)) \equiv \beta(\alpha)\beta'(\alpha) + r^2\sin^2\theta \rho_0(r) \chi'(\alpha).
\end{equation}
We do this by writing a discrete (finite-difference) version of this equation and solving the resulting system of algebraic equations. We discretize the space inside the star into a polar grid of $N_r$ regular intervals in the radial direction and $N_\theta$ in the angular direction, of length $\Delta r$ and $\Delta\theta$ each, respectively. On this grid, each point is labeled by a pair $(i,j)$, with coordinates $r_i = i\,\Delta r$ and $\theta_j = j\,\Delta\theta$. Points along the axis are those with $j=0$ (northern hemisphere) and $j = N_\theta$ (southern hemisphere) and points on the surface of the star correspond to $i = N_r$. Any quantity $q(r,\theta)$ evaluated on a point $(i,j)$ is labeled as $q_{i,j}\equiv q(r_i, \theta_j)$. The discretized GS equation reads
\begin{equation}\label{finite}
\frac{\alpha_{i+1,j} - 2\alpha_{i,j}+\alpha_{i-1,j}}{(\Delta r)^2} + \frac{1}{r_i^2}\frac{\alpha_{i,j+1}-2\alpha_{i,j}+\alpha_{i,j-1}}{(\Delta \theta)^2} - \frac{\cot\theta_j}{r_i^2}\frac{\alpha_{i,j+1}-\alpha_{i,j-1}}{2\,\Delta\theta} = -F(r_i,\theta_j, \alpha_{i,j}),
\end{equation}
where we use central difference for the derivatives. This algebraic equation is valid for $i=1,\hdots,(N_r-1)$, $j=1,\hdots,(N_\theta-1)$. The value of $\alpha$ at points lying on the axis ($\alpha_{i,0}$, $\alpha_{i,N_\theta}$ and $\alpha_{0,j}$) must remain fixed equal to zero, providing a boundary condition. Thus Eq. \eqref{finite} actually gives $(N_r-1)(N_\theta-1)$ algebraic equations for the $N_r(N_\theta - 1)$ unknown variables $\alpha_{i,j}$, $i=1,\hdots, N_r$, $j=1,\hdots, (N_\theta-1)$, so ($N_\theta-1$) additional equations are required in order to close the system, which are obtained from the boundary conditions at the stellar surface. Demanding continuity of the radial derivative of $\alpha$ gives the required number of equations,
\begin{equation}\label{surf}
\frac{\alpha_{N_r+1,j} -  \alpha_{N_r-1,j}}{2\Delta r} = -\frac{1}{R}\sum_{\ell=1}^{\ell_\text{max}}\frac{2\ell + 1}{2\ell + 2}\sin\theta_j P^1_\ell(\cos\theta_j)I_\ell \equiv  g_j
\end{equation}
(see Eq. \eqref{dadr}). Variables $\alpha_{N_r+1,j}$, standing for the value of $\alpha$ at points just outside the star, can be calculated by assuming that $\alpha$ outside the star is written by the multipolar expansion in Eq. \eqref{expoutside} (with $a_\ell = 0$), 
\begin{equation}
\alpha_{N_r+1,j} = \alpha_\text{out}(R+\Delta r,\theta_j),
\end{equation}
again, for $j=1,\hdots,(N_\theta-1)$, where we have introduced the parameter $\ell_\text{max}$ because we cannot perform the sum to infinity. The integral $I_\ell$ defined in Eq. \eqref{Iell}, and involved in expansions $g_j$ and $\alpha_\text{out}$, can be calculated through Simpson's rule and using the points at the surface, $\alpha_{N_r,j}$. This gives
\begin{equation}
I_\ell \approx \sum_{\substack{j=0\\j\,\text{even}}}^{(N_\theta-2)}\frac{\Delta \theta}{3}\left[P_\ell^1(\cos \theta_j)\,\alpha_{N_r,j} + 4\,P_\ell^1(\cos \theta_{j+1})\,\alpha_{N_r,j+1} + P_\ell^1(\cos \theta_{j+2})\,\alpha_{N_r,j+2}\right],
\end{equation}
so $N_\theta$ must be chosen as an \emph{even} natural number. Of course, using this form of the expansions, we are explicitly assuming continuity of $\alpha$ and its derivative with respect to $\theta$.

When introducing the explicit form of $I_\ell$ in terms of the $\alpha_{N_r,j}$'s, Eq. \eqref{surf} gives $(N_\theta-1)$ new independent equations relating part of the $N_r(N_\theta-1)$ unknowns on the grid, so a consistent solution of the algebraic equations may be carried out. The final system of equations to solve is
\begin{equation}
G_{i,j} \equiv \frac{\alpha_{i+1,j} - 2\alpha_{i,j}+\alpha_{i-1,j}}{(\Delta r)^2} + \frac{1}{r_i^2}\frac{\alpha_{i,j+1}- 2\alpha_{i,j}+\alpha_{i,j-1}}{(\Delta \theta)^2} - \frac{\cot\theta_j}{r_i^2}\frac{\alpha_{i,j+1} - \alpha_{i,j-1}}{2\,\Delta\theta} + F(r_i,\theta_j, \alpha_{i,j}) = 0,\label{Gij}
\end{equation}
for $i = 1,\hdots (N_r-1)$, $j = 1,\hdots,(N_\theta-1)$, and
\begin{equation}
G_{N_r,j} \equiv \frac{\alpha_{N_r+1,j} -  \alpha_{N_r-1,j}}{2\Delta r} - g_j = 0,\label{GNrj}
\end{equation}
for $j=1,\hdots,(N_\theta - 1)$, where
\begin{equation}
\alpha_{N_r+1,j} = \frac{\Delta\theta}{6}\sin\theta_j\sum_{\substack{\tilde\jmath = 0\\\tilde\jmath\,\text{even}}}^{(N_\theta - 2)}\left[\alpha_{N_r,\tilde\jmath}\, T_{j,\tilde\jmath} + 4\,\alpha_{N_r,\tilde\jmath +1}\,T_{j,\tilde\jmath+1} + \alpha_{N_r,\tilde\jmath + 2}\, T_{j,\tilde\jmath+2}\right]
\end{equation}
and
\begin{equation}
g_j = -\frac{\Delta\theta}{6R}\sin\theta_j \sum_{\substack{\tilde\jmath = 0\\ \tilde\jmath\,\text{even}}}^{(N_\theta - 2)}\left[\alpha_{N_r,\tilde\jmath}\, U_{j,\tilde\jmath} + 4\,\alpha_{N_r,\tilde\jmath +1}\,U_{j,\tilde\jmath+1} + \alpha_{N_r,\tilde\jmath + 2}\, U_{j,\tilde\jmath+2}\right].
\end{equation}
Here, $T_{j,\tilde\jmath}$ and $U_{j,\tilde\jmath}$ stand for 
\begin{equation}
T_{j,\tilde\jmath} \equiv \sum_{\ell = 0}^{\ell_\text{max}}\left(\frac{R}{R+\Delta r}\right)^\ell \frac{2\ell + 1}{\ell(\ell + 1)}\,P_\ell^1(\cos\theta_j)\,P_\ell^1(\cos\theta_{\tilde\jmath})\qquad\text{and}\qquad U_{j,\tilde\jmath} \equiv \sum_{\ell = 0}^{\ell_\text{max}}\frac{2\ell + 1}{\ell + 1}\,P_\ell^1(\cos\theta_j)\,P_\ell^1(\cos\theta_{\tilde\jmath}).
\end{equation}
It is important to notice that, despite the number of equations and unknowns are equal, the system is still nonlinear as $F(r,\theta,\alpha)$ is in general nonlinear in $\alpha$. We developed a code in Mathematica 9.0 to solve the nonlinear, algebraic system of equations in \eqref{Gij} and \eqref{GNrj}. Our code is based on a generalized Newton's method of the form
\begin{equation}\label{newton}
J_G(\alpha_{i,j}^{(n)})(\alpha_{i,j}^{(n+1)} - \alpha_{i,j}^{(n)}) + G_{i,j}^{(n)} = 0,
\end{equation}
where $G_{i,j} = 0$, $i=1,\hdots, N_r$, $j=1,\hdots,(N_\theta - 1)$, is the system of (nonlinear) equations to be solved, $J_G(\alpha_{i,j})$ is the Jacobian of the system $G$ and $n$ is an index denoting the $n-$th iteration. The aim is to solve for the $(n+1)$th step given the $n$th one, starting from an initial seed $\alpha_{i,j}^{(0)}$ until certain convergence criterion is achieved, namely, 
\begin{equation}
|\alpha_{i,j}^{(n+1)} - \alpha_{i,j}^{(n)}| \leq \epsilon|\alpha_{i,j}^{(n)}|,
\end{equation} 
for all $i$ and $j$, where $\epsilon$ is an arbitrary small number. Notice that, in doing so, Eq. \eqref{newton} is a \emph{linear} system of equations, easily solved by standard methods of linear algebra when the Jacobian is non-singular. Once the system $G$ is solved, we can compute back the coefficients $b_\ell$ that determine $\alpha$ outside the star, in this way obtaining $\alpha$ everywhere. \\

\end{document}